\documentclass[12pt,preprint]{aastex}

\shorttitle{The Sextet Arcs towards A1689}
\shortauthors{Frye et al.}

\begin{document}


\title{The Sextet Arcs:  a Strongly Lensed Lyman Break Galaxy in the ACS Spectroscopic Galaxy Survey towards Abell 1689}

 
\author{Brenda L. Frye\altaffilmark{1}$^,$\altaffilmark{2}, Dan Coe\altaffilmark{3},
David V. Bowen\altaffilmark{4}, 
Narciso Ben{\'  \i}tez\altaffilmark{3}, Tom Broadhurst\altaffilmark{5}, Puragra Guhathakurta\altaffilmark{6},
Garth Illingworth\altaffilmark{6}, Felipe Menanteau\altaffilmark{7},
Keren Sharon\altaffilmark{5}, Robert Lupton\altaffilmark{4}, Georges Meylan\altaffilmark{8}, 
Kerry Zekser\altaffilmark{9},Gerhardt Meurer\altaffilmark{9}, and Mairead Hurley\altaffilmark{1}}






\altaffiltext{1}{Department of Physical Sciences, Dublin City University, Glasnevin, Dublin 9, Ireland; brenda.frye@dcu.ie}
\altaffiltext{2}{Council on Science \& Technology Fellow, Princeton University}
\altaffiltext{3}{Instituto de Astrofisica de Andaluc{\'i}a (CSIC), C/Camino Bajo de Hu{\'e}tor, 24, Granada, 18008, Spain}
\altaffiltext{4}{Department of Astrophysical Sciences, Peyton Hall, Princeton University, Princeton, NJ  08540}
\altaffiltext{5}{School of Physics and Astronomy, Tel Aviv University, Tel Aviv 69988, Israel}
\altaffiltext{6}{UCO/Lick Observatory, University of California, Santa Cruz, CA  95064}
\altaffiltext{7}{Department of Physics and Astronomy, Rutgers the State University of NJ, Piscataway, NJ 08854}
\altaffiltext{8}{Laboratoire d'Astrophysique, Ecole Polytechnique F\'ed\'erale de Lausanne (EPFL) Observatoire, CH-1290, Sauverny, Suisse}
\altaffiltext{9}{Physics and Astronomy Department, Johns Hopkins University, Baltimore, MD}

\begin{abstract}

We present results of the HST Advanced Camera for Surveys
spectroscopic ground-based redshift survey in the field of A1689.  We
measure 98 redshifts, increasing the number of spectroscopically
confirmed objects by sixfold.  We present two spectra from this
catalog of the Sextet Arcs, images which arise from a strongly-lensed
Lyman Break Galaxy (LBG) at a redshift of $z$=3.038.  Gravitational
lensing by the cluster magnifies its flux by a
factor of $\sim$16 and produces six separate images with a total
$r$-band magnitude of $r_{625}=21.7$.  The two spectra, each of which
represents emission from different regions of the LBG, show H~I and
interstellar metal absorption lines at the systemic redshift.
Significant variations are seen in Ly-$\alpha$ profile across a
single galaxy, ranging from strong absorption to a combination of
emission plus absorption.  A spectrum of a third image close to the
brightest arc shows Ly-$\alpha$ emission at the same redshift
as the LBG, arising from either another spatially distinct region of
the galaxy, or from a companion galaxy close to the LBG.  Taken
as a group, the Ly-$\alpha$ equivalent width in these three
spectra decreases with increasing equivalent width of the
strongest interstellar absorption lines.  We discuss how these
variations can be used to understand the physical conditions in the
LBG.  Intrinsically, this LBG is faint, $\sim$0.1$L^*$, and forming
stars at a modest rate, $\sim$4 $M_{\odot}~{\rm yr}^{-1}$.  We also
detect absorption line systems toward the Sextet Arcs at $z$=2.873 and
$z$=2.534.  The latter system is seen across two of our spectra.

\end{abstract}



\keywords{galaxies: clusters: general --- galaxies: clusters: individual (A1689)---galaxies: high-redshift---gravitational lensing---techniques: spectroscopic---methods: data analysis}


\section{Introduction}

From a certain observational standpoint, our understanding of
Lyman-break galaxies (LBGs) is akin to that of the Milky Way nearly
half a century ago.  The discovery of the connection between stellar
abundances and kinematics to the history of galaxy evolution made for
a significant step forward in the growing understanding of galaxies
\citep{Freeman:02,Eggen:62}.

For LBGs, the large leap forward came with the discovery of nearly
one-thousand galaxies at $z\sim$3 initially targeted via color
selection for the Lyman-series break inherent in their spectra
\citep{Steidel:03}.  The general properties of this galaxy population
have been investigated in detail \citep{Adelberger:03, Adelberger:98,
Steidel:98}.  To examine the intrinsic properties of LBGs as a group,
\citet{Shapley:03} produced a composite spectrum from over 800
individual LBG spectra.  They found that the combined absorption plus
emission rest equivalent width (REW) of Ly-$\alpha$ depended on four
primary spectral characteristics: for increasing Ly-$\alpha$ REW the
total REW of the low ionization interstellar lines decreased, the rest
frame UV spectrum became bluer, the velocity offset between
Ly-$\alpha$ and the mean redshift of the interstellar lines decreased,
and the star formation rate went down.  They explained these effects
as a combination of incomplete coverage of the UV continuum light by
gas and dust, and a range of velocities over which gas is absorbed.

The coaddition of many hundred LBG spectra by \citet{Shapley:03} was
necessary because individual LBGs are faint, and the spectral
properties of individual LBGs are hard to record at high spectral
resolution and signal-to-noise. A notable exception is the
strongly-lensed LBG MS1512$-$cB58 (hereafter cB58) at $z$=2.73
\citep{Yee:96}.  Magnification by the cluster MS1512$+$36 increases
the flux of the LBG to $r_{625}=20.4$, making it possible to observe
this galaxy at a resolution high enough to obtain accurate abundances
from interstellar absorption lines.  \citet{Pettini:02, Pettini:00}
found the gas outflowing from cB58 to be highly enriched in
$\alpha$-elements (from Type~II supernovae), with abundances of O, Mg,
Si, P and S all $\sim0.4$ times the solar value.  At the same time,
the N and Fe-peak elements of Mn, Fe and Ni were underabundant by a
factor of three. Such a pattern implied that the outflowing gas had
been enriched for only $\sim 300$~Myr.

The class of strongly-lensed LBGs (hereafter SLLBGs), of which cB58 is
a member, is characterized by its enhanced flux.  Ordered by
increasing $r_{625}$-band flux, the recently discovered `8 o'clock
arc' is the brightest with a total magnitude of $r_{625}$=19.22
\citep{Allam:06}.  It is followed by LBG J2135.2-0102 the `Cosmic Eye'
\citep{Smail:06}, cB58, A2218-384 \citep{Ebbels:96}, the Sextet Arcs
(this paper), 1E0657-56-A \citep{Mehlert:01}, and Q0000-D6
\citep{Giallongo:02}.  
Note there is also one bright strongly-lensed galaxy known at lower
redshift, $z$=1.9, but the Ly-series break is not redshifted into the
optical bands and so it is not included here
\citep{Lemoine-Busserolle:03}.  They are brighter than the majority of
field LBGs, making medium spectral resolution observations feasible.
Additionally, the five brightest SLLBGs, $r_{625}<22.5$ (including the
one presented in this paper), are also extended on the sky, enabling
spatially-resolved spectroscopy of high redshift galaxies. This has
already been achieved for two fainter, higher redshift objects: the
quadruply-lensed giant arc in A2390 at $z$=4.040
\citep{Frye:97,Bunker:98}, and a galaxy at $z\sim5$
\citep{Swinbank:07}.

While detailed studies of this class are in progress, there are
already clues that the intrinsic properties of luminosity and
extinction for at least some of the SLLBGs are different from the
general LBG population.  The brightest SLLBG, the 8 o-clock arc, has
an estimated unlensed magnitude that is brighter than the general LBG
population by four magnitudes in $r_{625}$.  In another recently
discovered SLLBG, the so-called Cosmic Eye arc, LBG J2135.2-0101, the
extinction is high enough to place it on the borderline of the reddening
selection criterion for the \citet{Steidel:03} sample.

We present here spectra of a new SLLBG at $z$=3.038 whose lensed
images we designate as the ``Sextet Arcs."  The Sextet Arcs consist of
six separate images towards the massive lensing cluster A1689
($z$=0.187) with a total magnitude integrated over all images of
$r_{625}=21.7$.  This bright LBG was first identified by
\citet{Broadhurst:05} as a part of the strong lensing study of A1689
and is relatively rare in being a multiply-imaged galaxy with a
reported spectroscopic redshift.  Only three of the $\sim$30 multiply
imaged galaxies have confirming spectra, and thus this system of arcs
has featured prominently in several recent strong lensing analyses
used to derive dark matter mass distributions in this well studied
cluster  \citep{Leonard:07,Limousin:06, Zekser:06, Halkola:06,
Broadhurst:05}.

Most of these data were taken for the HST {\it Advanced Camera for
Surveys} (ACS) spectroscopic ground-based redshift survey.  Some data
were acquired prior to the start of this GTO program on A1689.  We
present a redshift catalog of nearly 100 objects in the field of
A1689, with nearly three-fourths of them being new arclets with
$z>$0.23.  Until this paper, only the redshifts of 10 objects in the
background of A1689 have appeared in the literature \citep{Broadhurst:05, Frye:02}.

This paper is organized as follows: We summarize the imaging and
spectroscopic observations in \S2, and outline the custom-built
reductions and analysis in \S3.  We show the spectra for the Sextet
Arcs in \S4, and the spectra for the spatially-resolved intervening
absorption system in \S5.  In \S6 we discuss the
variations of spectral characteristics seen along the Sextet Arcs and
the $z$=2.534 absorption system and how these results can be used to
further constrain the lens model.  We also measure the intrinsic
properties of star formation rate and intrinsic luminosity.  Finally
in \S7 we give the summary and conclusion.  The results of our
spectroscopic catalog, and comparison with the literature, is given in
the Appendix.  Throughout this paper we assume a cosmology with 
$H_0=70$ km s$^{-1}$ Mpc$^{-1}$, $\Omega_{m,0} = 0.3$, $\Omega_{\Lambda,0} = 0.7$.

\section{Observations}

\subsection{Imaging}

Abell 1689 was observed in June, 2002 as part of HST ACS GTO time.
Deep exposures were taken in the $g_{475}$, $r_{625}$, $i_{775}$, and
$z_{850}$ passbands, reaching limiting magnitudes of 27.5, 27.2, 27.2
and 26.7 respectively.  Details of the ACS observations can be found
in \citet{Broadhurst:05}.  To calculate robust photometric redshifts,
we obtained $UBVRIZJHK$ ground-based images.  The $U$-band image was
acquired from the DuPont Telescope at Las Campanas and was taken in
conditions of $0\farcs93$ seeing, as measured from the PSF off the
images.  The $B$-band image was acquired from the Nordic Optical
Telescope (NOT) at La Palma, from which we measure $1\farcs13$ seeing.
Broadband $V$, $R$, $I$ images, and a narrow-band $Z$ band image taken
with the NB9148 filter were obtained in March 1999 with LRIS on Keck 2
taken under conditions of $0\farcs81$, $0\farcs74$, $0\farcs68$, and
$0\farcs91$ seeing, and limiting magnitudes of 27.17, 26.96, 26.41,
and 24.81 respectively for the four filters.  Finally, the $J$, $H$,
and $Ks$ images were obtained with Son of Isaac (SOFI) on the ESO New
Technology Telescope (NTT) taken under conditions of $0\farcs87$,
$0\farcs99$, and $0\farcs82$ seeing to limiting magnitudes of 25.12,
24.22, and 24.06.  A complementary $Ks$-band Infrared Spectrometer and
Array Camera (ISAAC) VLT image taken under conditions of $0\farcs95$
seeing was also analyzed.

\subsection{Imaging Results}
\begin{figure}
\plotone{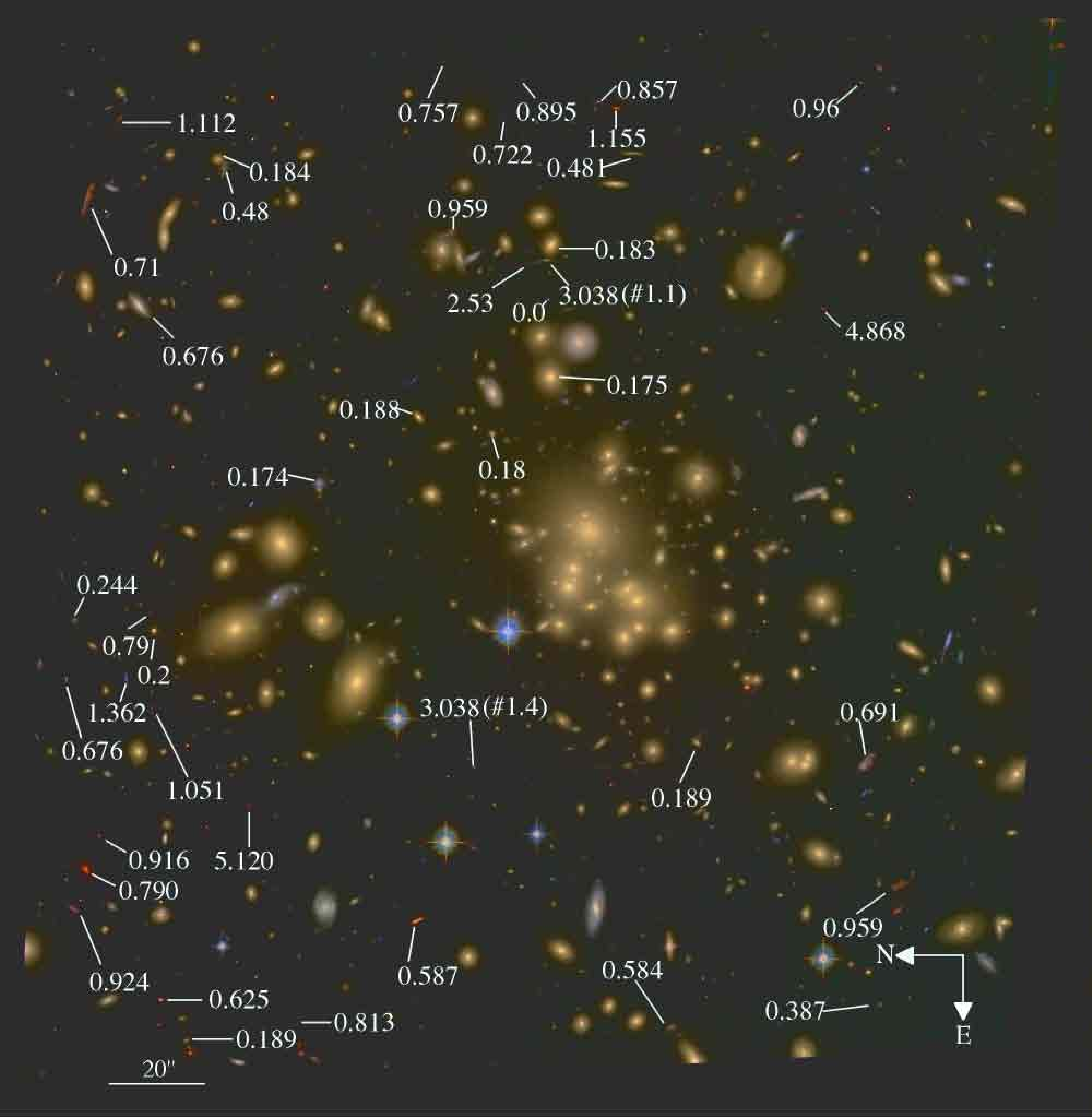}
\caption{Deep HST ACS $gri$ true color image of the central portion of Abell 1689.  
It was made  
using the Sloan Digital Sky Survey pipeline photo \citep{Lupton:01}.  
All new spectroscopic redshifts in this field are indicated, with two of the six
separate images of the Sextet Arcs labeled  (\#1.1 and \#1.4 at $z$=3.038). 
The Sextet Arcs are a strongly-lensed LBG with a total magnitude integrated over all the arcs
of $r_{625} = 21.7$, bright enough to do high spectral resolution followup work.
\label{fig2}}
\end{figure}

The ACS $gri$ true color image of the central portion of Abell 1689 is
shown in Figure~\ref{fig2}.  This image was made using the Sloan
Digital Sky Survey pipeline photo \citep{Lupton:01}.  The asinh
stretch was applied, as described in \citet{Lupton:04}; it is fast,
and qualitatively similar to the ACS image presented by
\citet{Broadhurst:05}.  Cluster members appear as extended yellow
objects in the center and left of center, and the objects from our
sample in common with our larger Keck LRIS field are marked.
Additionally, two images of the six Sextet Arcs at $z=3.038$ are
circled.

The Sextet Arcs were identified by \citet{Broadhurst:05} to be one
object lensed into six separate arcs.  The arcs are labeled \#1.1
through \#1.6 and each of them are shown in our HST ACS $r_{625}$-band
data, (Figure~\ref{fig7}).  The image in the upper left panel of this
6-panel image is a fold arc comprising two unseparated images but
taken to be one image, named \#1.1.  Image \#1.1 has an estimated
magnified image area of $\sim$16$\times$, as measured from the
delensed images in \citet{Broadhurst:05}.  Images \#1.1 and \#1.2
subtend a length of $\sim$6 arcseconds, similar in apparent size to
the other known strongly-lensed LBGs.  The five images comprising
source \#2, which we designate as the `Quintet Arcs,' are also shown.
Table~\ref{table3} gives the positions of all images of the  Sextet Arcs and Quintet Arcs in
ACS $g_{475}$, $r_{625}$, and $i_{775}$-band magnitudes.  

We recount the argument in \citet{Broadhurst:05} to show that the six
separate images comprising the Sextet Arcs all belong to one source
galaxy.  First, each image of the Sextet Arcs, except for the central
demagnified one, has the same elongated morphology.  Each image of the
Sextet Arcs is also always accompanied by a neighboring image which
has a distinctive core-plus-halo morphology.  These neighboring
images, with an average angular separation of $\sim$2 arcsec depending
on the differing magnifications for each image, are the Quintet Arcs
(Source \#2).  The image pairs, comprising one image each of the
Sextet Arcs (Source \#1) and Quintet Arcs (Source \#2), appear with
strikingly-similar morphologies at multiple locations despite being
stretched, rotated, and parity-flipped, thus allowing for their
confident identification by-eye as multiply-lensed
(Figure~\ref{fig7}).  From this initial identification a mass model
was constructed and used to predict and verify the positions of other
counterarcs, including the central demagnified image.  Photometric
redshifts were then calculated and found to match the data.  Finally,
spectra were taken of two of the images which yielded the same
redshift, $z=3.038$.

\subsection{Spectroscopy}
A summary of the ground-based spectroscopic observations is given in
Table~\ref{table2}.  Observations were carried out at the Keck
Observatory, the Very Large Telescope (VLT), and Las Campanas
Magellan-Clay Observatory.  Targets were color-selected in Keck LRIS
bands $V_{AB}-I_{AB}$ to be behind the cluster, and the limiting
$I_{AB}$ magnitude selected to suit the allocated observing time and
instrument.  In total, spectra were obtained for 255 objects.  Of
those, 98 objects with secure spectroscopic redshifts are presented
here (see Tables~\ref{table1} and \ref{table3}).

\subsubsection{Keck Observations}
  
Nearly three-fourths of the data were obtained from Keck LRIS over a
course of five observing runs, from April 1996 to March 2003.  Seven
multislit masks were used with either the 300 lines/mm grating (blazed
at a central wavelength of 5000 \AA), or the 400 lines/mm grating
(blazed at 8500 \AA).  The observed resolution varied slightly over
the observing runs, with the lower dispersion grating providing a
resolution of 12 \AA \ FWHM and the higher one, 9 \AA \ FWHM,
determined from the 6300 \AA \ skyline.  Typical exposure times for
both setups were 6 $\times$ 1200 sec with 1\farcs0 dithers between
exposures to correct for fringing and uneven illumination along the
slits.  The slit widths were chosen to be from 0.8\arcsec $-$1\arcsec,
and the conditions ranged from $\sim 0.65$\arcsec $-$ 0\farcs85
seeing. The spectrum of the brightest image of the Sextet Arcs, \#1.1,
was obtained in June, 1997 with the 300 line/mm grating in 2.8 hours
of integration (see Figure~\ref{fig7}).

\subsubsection{VLT Observations}
Spectroscopic data were obtained at VLT on FORS2 in June and July 2001
in service mode.  One multislit mask was used with the 300 lines/mm
grism, providing 12 \AA \ resolution at 6300 \AA.  Exposure times were
33 $\times$ 1200 sec with 1\farcs0 dithers between exposures.  The
slitwidths were 1\arcsec.  Objects fainter than $I_{AB} =23$ were
targeted, and the results are given in Table~\ref{table2}.  One high
redshift galaxy was discovered, an emission line object at $z$=4.705
and $I_{AB}$=25.3 (see Table~\ref{table3}).  Also additional data on
the bright triply-lensed system at $z$=4.868 was obtained and will
appear in an upcoming paper (see Frye et al. 2002 for the Keck LRIS
spectrum of this object).

\subsubsection{Magellan Observations}
One-fourth of the survey was carried out at Magellan Observatory in
2003 May on the low-dispersion survey spectrograph (LDSS2).  Five
multislit masks were used with the 300 lines/mm medium-blue grism
blazed at 5500 \AA.  The grism provided 16 \AA \ resolution at 6700
\AA.  Exposure times varied from 2 $\times$ 3600 sec to 14 $\times$
3600 sec, with 1\arcsec \ dithers after every two exposures.  Spectra
for arclets \#1.4 and \#2.1 were obtained in a total of 8 h and 11 h
respectively.


\section{Data Reduction}
\subsection{Spectra}
We have custom-built a spectroscopic reduction package for this
project.  The aim is to maximize the signal to noise of background
limited data without resampling the pixels, so that groups of pixels
carrying faint continuum signal have every chance of being detected as
a coherent pattern in the final reduced image.  It can operate on full
2d fits frames directly, outputting reduced 2d frames.  It is unique
in that it can follow the object signal over the full dispersion range
even if the object signal drops off, by following the curvature of the
similarly-distorted spectrum box edges.  Also it can correct for mask
defects such as the complex spatial slit profile, and can accommodate
subpixel shifts between the flatfield and the data frames by modeling
the slit profile with a dispersed flatfield.  The details of the
reduction pipeline can be found in \citet{Frye:02}.

The spectral data for image \#1.1 of the Sextet arc was fluxed by
taking the spectrum of an early-type cluster member at $z$=0.187,
observed simultaneously through the multislit mask, and comparing it
to the standard empirical E/SO spectrum of \citet{Kennicutt:92}.  The
spectral data for images \#1.4 and \#2.1 from Magellan were taken at
low $\sim$20 \AA \ resolution.  The data were flux calibrated with the
standard star LTT3218.  A sensitivity function was created using the
IRAF task sensfunc, which was then scaled to the spectrum and divided
into the data.

\subsection{Photometry}
Objects within the ACS FOV were detected in a
$g\arcmin$+$r\arcmin$+$i\arcmin$+$z\arcmin$ ``galaxy-subtracted''
image using SExtractor \citep{Bertin:96}.  The galaxy subtraction
involved carefully modeling the A1689 cluster galaxies, and
subtracting their light from each image.  This helps both to detect
extra galaxy images (especially near the center of the cluster) and to
improve the photometry of the faint, low-surface brightness background
galaxies.  This meticulous process has not been extended to our
ground-based images.  Bayesian photometric redshifts were obtained
based on this galaxy-subtracted These redshift estimates provided a
means of confirming the thirty sets of multiply-imaged objects
identified in the field \citep{Broadhurst:05}.  These images are very
faint, $23.4<i<28$, making spectroscopy unfeasible for all $>$100
images.

To include galaxies within a larger FOV, objects were also detected in
the deepest Keck image (the $I$-band).  Aperture-matched PSF-corrected
photometry was then obtained across the 14-filter ACS and ground-based
image set using techniques since made available in the ColorPro
software package \citep{Coe:06}.  A separate Bayesian photometric
redshift catalog was obtained based on these magnitudes.  A good
general agreement is found between measured redshifts for objects
detected in both catalogs.  For objects within the ACS FOV, we report
photometric redshifts as measured in the above-mentioned ACS catalog,
while for objects outside the ACS FOV, we quote results from this
Keck-based catalog.  More details will be published along with the
full photometric catalogs in an upcoming paper \citep{Coe:07}.

\section{Source \#1:  The Sextet Arcs}
\subsection{Image \#1.1}
The spectrum of image \#1.1 was optimally-extracted and is shown in
Figure~\ref{fig_dvb}.  Several strong low ionization absorption lines
are detected.  They are Si II $\lambda$1260, O~I~$\lambda 1302$ +
Si~II~$\lambda1304$ (where we take the mean wavelength of these
blended lines to be 1303~\AA), C~II $\lambda$1334, SiIV
$\lambda\lambda$1393, 1402, Si II $\lambda$1527, Fe II $\lambda$1608,
and Al II $\lambda$1671.  The rest equivalent widths of these lines
are listed in Table~\ref{table3}, including their 1-$\sigma$ errors.
A fit to the line centroids of the four strongest interstellar
absorption lines, Si II $\lambda$1260, O~I $+$ Si~II $\lambda_{mean}=
1303$, C~II $\lambda$1334, and Si II $\lambda$1527, yields an
absorption line redshift of $z$=3.041.

The Ly-$\alpha$ absorption line profile shows considerable structure,
 some of which appears to be caused by the superposition of a weak
 emission line at 4907~\AA. Although this feature may simply be an
 artifact of complex multicomponent Ly-$\alpha$ absorption lines close
 to the redshift of Source \#1, we find a similar, unambiguously
 identified Ly-$\alpha$ emission line in the spectrum of image \#1.4
 at an identical redshift (see \S5.2). For this reason, we believe the
 weak feature near the deepest part of the Ly-$\alpha$ absorption line
 profile in the spectrum of image \#1.1 is indeed a Ly-$\alpha$
 emission line.  As there are no stellar photospheric features, we
 cannot measure the systemic redshift of the SLLBG directly.  We
 obtain the velocity offset correction by following the prescription
 in \citet[their equation (5)]{Adelberger:03}.  Based on a mean
 velocity difference of -75 km s$^{-1 }$ between Ly-$\alpha$ emission
 and the low ionization interstellar lines, we calculate a velocity
 offset of $240$ km s$^{-1}$ smaller than the nebular redshift, or
 $z_{sys}$=$3.038\pm0.003$.

The Ly-$\alpha$ absorption is wide, and has a shape reminiscent of a
Damped Ly-$\alpha$ (DLA) line.  However, the flux of the line never
reaches zero, as would be expected for a DLA observed at the
resolution of our data.  As discussed above, this is probably because
the absorption is filled in with Ly-$\alpha$ emission.  Apart from
this contamination, the underlying absorption must arise from some
combination of high and low H~I column density clouds overlapping in
velocity.  At the resolution of our data, however, we cannot untangle
the velocity structure of the absorption.  On the other hand, we can
at least set an upper limit to the H~I column density, $N${H~I},
assuming that the absorption comes from a single high-$N$(H~I) cloud.
We can set the absorption redshift to be the same as that of the SiII
$\lambda$1260 line, which is the only metal line not blended with
metal lines from lower redshift absorption line systems.  The redshift
of this line is $z$=3.038 (the same as the Ly-$\alpha$ emission) must
therefore reflect the redshift of the bulk of the H~I gas.  The upper
limit to $N$(H~I) can then be determined by fitting the red wing of
the Ly-$\alpha$ absorption between $\approx$1200-1250 \AA.  This
procedure is not straightforward, since not only do we not know the
full velocity extent of the contaminating Ly-$\alpha$ emission, and
the wing may well be contaminated by SiIV $\lambda \lambda$ 1393, 1402
at $z$=2.534 (see \S6.  Using the available continuum, however, it
seems likely that the H I column density is log $N$(H~I) $\le 21.5$.
We also find that at $z$=3.038, the Ly-$\beta$ line expected from the
Ly-$\alpha$ absorption does not line-up well with the feature
identified as Ly-$\beta$ at 4122 \AA.  It thus seems likely that the
Ly-$\beta$ absorption line is also filled in with Ly-$\beta$ emission,
and cannot be used to better constrain $N$(H~I).

The low ionization absorption lines are all strong, and reminiscent of
the SLLBG MS1512-cB58 \citep{Pettini:02}.  A comparison of rest
equivalent widths of all the lines that are in common with those seen
towards cB58 shows that the absorption in the spectrum of image \#1.1
is weaker.  Only C~II $\lambda$ 1334 is of equal strength, but this is
due to a chance blend with two other intervening absorption lines,
SiIV $\lambda\lambda$1394, 1402 at $z$=2.873, and SiII $\lambda$1527
at $z$=2.534.  As is true for cB58, the metal lines in our data are
also likely saturated.  For example, the ratio of the rest equivalent
width of SiII $\lambda$1260 to SiII $\lambda$1527 should be $\sim$6
for optically thin lines \citep{Morton:03}.  We measure the ratio of
rest equivalent widths to be $W_0(1260)/W_0(1527)=1.9 \pm 0.73$, as
compared to 0.95 for the composite LBG spectrum \citep{Shapley:03}.
As for SiIV $\lambda\lambda$1393, 1402, while
we do detect both lines, they are blended with AlII $\lambda$ 1608 at $z$=2.534 and so
are of no use other than redshift confirmation.  
Note at our resolution we do not detect any weak absorption
features which fall on the linear part of the curve of growth.  

The spectrum of image \#1.1 appears to show little evidence for gas
motions.  The velocity difference between the line centroid of
Ly-$\alpha$ emission and the mean of the interstellar absorption lines
is $\Delta v =-75 \pm150$ km~s$^{-1}$.  Gas with such a velocity
offset that is consistent with a static medium or even a blueshift is
observed in only $\sim$5\% of cases in the general LBG population
\citep{Shapley:03}.  The theoretical prediction from Monte Carlo
simulations of $z\sim3$ galaxies is for Ly-$\alpha$ to be redshifted
from the stellar redshift by twice the velocity shift of the
interstellar lines from Ly-$\alpha$ \citep{Verhamme:06}.  For our data
Ly-$\alpha$ is blueshifted from the interstellar lines by $\Delta v =
150$ km~s$^{-1}$, and Ly-$\alpha$ is also blueshifted from the stellar
redshift, by $\Delta v_{sys}= 200$ km~s$^{-1}$.  At the same time,
the absorption line profile of Si~IV $\lambda$1394 shows broad blueshifted absorption indicative of gas outflow.
We require higher resolution data
to address the somewhat unusual kinematics of this galaxy. 

To measure the extinction in this spectrum, we fit our multiwavelength
photometry to the starburst SED templates of \citet{Calzetti:94}.  Six
starburst galaxy templates are provided with various levels of
extinction.  To these we add interpolations of adjacent templates
until a best fit is found.  We measure $E(B-V)=0.47$, and consider
this value to be an upper limit owing to an unknown level of light
contamination from a nearby cluster elliptical at $z=0.183$.

In addition to the absorption at $z$=3.038, there are also two other
intervening absorption systems, both of which are marked in
Figure~\ref{fig_dvb}.  The first is at $z$=2.873.  It shows strong
Ly-$\alpha$ absorption, O I $\lambda$1302 $+$ Si II $\lambda$1304,
C~II $\lambda$1334, SiIV $\lambda$1393 (blended with C~II
$\lambda$1334 at $z$=3.038 and Si~II $\lambda$1527 at $z$=2.534), and
Si II $\lambda$1527.  The second absorption system is at $z$=2.534.
It shows O I $\lambda$1302 $+$ Si II $\lambda$1304, $\lambda_{mean}=
1303$, SiIV $\lambda\lambda$1393, 1402, Si II $\lambda$1527, Fe II
$\lambda$1608, and Al II $\lambda$1671.  Interestingly, we observed
the absorption system at $z$=2.534 also in another spectrum taken only
2 arcsec away on the sky, which will be discussed in \S5.

\subsection{Image \#1.4}
Image \#1.1 discussed above and image \#1.4 are both images of the
same source.  In Figure~\ref{fig_1d} we show the spectrum of image
\#1.4.  It shows strong Ly-$\alpha$ emission and Ly-series decrement
(Figure~\ref{fig_1d}).  We see indications of gas motions in the
linewidth of the Ly-$\alpha$ line as indicated by a large deconvolved
FWHM = 1700 km s$^{-1}$.  Further, the P-Cygni type profile of
Ly-$\alpha$ is indicative of large scale gas outflow, and is in
contrast to the Ly-$\alpha$ absorption seen in image \#1.1
(Figure~\ref{fig_1d}).  The implications of the variations in
Ly-$\alpha$ seen across a {\it single} galaxy will be discussed in
\S7.

We calculate the extinction in image \#1.4 to be $E(B-V)=0.10$.  It
was calculated in the same way as that for image \#1.1, discussed
above.  Given the small size of this image compared to the slitwidth
of the multislit mask, we are confident that we are capturing all of
the stellar component of this LBG in the spectrum of image \#1.4.
This is in contrast to image \#1.1, for which the spectrum records
only 1 arcsec of the giant ($>$5.5 arcsec long) arc.  For this reason
we adopt our measured value of the extinction of $E(B-V)$=0.10 as the
best value for the Sextet Arcs.  The value is similar to the values
measured by \citet{Shapley:03} in their composite LBG spectra.


\section{Source \#2:  The Quintet Arcs}
The Quintet Arcs (our designation for the five images of Source \#2
identified by Broadhurst et al. 2005), are shown in Figure~\ref{fig7}.
They are spatially-resolved and characterized by a two component
compact bright core plus extended halo.  This morphology is
recognizable from image to image along the critical curve of the
cluster despite being stretched, rotated, and parity-flipped.  The
Quintet Arcs subtend an average lensed separation from the Sextet Arcs
of 2 arcsec, with the variation in angular separation between the
images accounted for by the differing magnifications along this second
strongly-lensed object in this paper.

\subsection{Image \#2.1}

The spectrum for image \#2.1 is shown in Figure~\ref{fig_1d}.  Unlike
the spectra for either image \#1.1 or \#1.4, Ly-$\alpha$ is seen in
emission only, with no corresponding Ly-$\alpha$ absorption. The low
ionization lines of SiII $\lambda$ 1260 and C~II $\lambda$1334 (Si~II
$\lambda$1527 at $z$=2.534 and SiIV $\lambda\lambda$1393, 1402 at
$z$=2.87).  We see indications of gas motions in the linewidths of
this spectrum, reporting a deconvolved FWHM for Ly-$\alpha$ of 1350 km
s$^{-1}$.  The total profile of emission plus absorption for
Ly-$\alpha$ is symmetric, in contrast both to the absorption seen in
the spectrum of image \#1.1 and the P-Cygni type profile seen towards
image \#1.4.  We measure $E(B-V)=0.17$, similar to image \#1.4.

There is an intervening absorption system towards this object at
$z$=2.534 (Figure~\ref{fig_1d}).  Several low ionization lines are
detected, including Si II $\lambda$1260, O I $\lambda$1302 $+$ Si II
$\lambda$1304, C~II $\lambda$1334, Si II $\lambda$1527, C IV
$\lambda\lambda$1448, 1550, and Fe II $\lambda$1609.  Ly-$\alpha$
absorption clearly reaches the zero-level of the continuum, and is
possibly damped.

\subsection{Source Correspondence and Redshift Determination for Image \#2.1}

The slit used to obtain a spectrum of image \#2.1 unexpectedly
produced a spectrum with precisely the same redshift as Sextet Arc
image \#1.1, including a strong emission feature at the expected
position of Ly-$\alpha$ at $z$=3.038.  However, we think it unlikely
that this is the true redshift of \#2.1.  We searched for the same
Ly-$\alpha$ emission feature in our other spectra of Source \#2, but
only our 3 h LDSS2 Magellan spectrum of image \#2.2 covered the
correct wavelength range.  The observed equivalent width of
Ly-$\alpha$ in the spectrum of image \#2.1 is measured to be
$W_{obs}=20$ \AA.  A comparison measurement made of a typical noise
feature near the expected position of Ly-$\alpha$ in the spectrum of
image \#2.2 gives $W_{obs}=40$ \AA.  It is thus unlikely that we would
be able to detect the same Ly-$\alpha$ seen in the spectrum of image
\#1.1 in the available spectrum of image \#2.2.

The idea that there may be significant contamination from image \#1.1
at the position of image \#2.1 was first put forth by
\citet{Broadhurst:05}, given that both images are found near the
critical curve and stretched significantly on the sky.
Figure~\ref{fig10} shows the observational setup used to record the
spectra of images \#1.1 and \#2.1 using a one arcsec slit centered on
image \#2.1, and a one arcsec slit centered on the brightest part of
image \#1.1.  This separation is only two arcsec on the sky, which may
not be enough to clear it of stray light from \#1.1.  In fact, the
Ly-$\alpha$ emission peak in the spectrum of image \#2.1 is found to
be separated in velocity by only $+40$ km s$^{-1}$ from the
Ly-$\alpha$ emission from image \#1.1.

As discussed in \S6.2, there is a strong intervening absorption system
at $z$=2.534. While it is not possible to measure a column density at
this resolution, the absorption is so broad, 4000 km s$^{-1}$, that it
is probably a damped Ly-$\alpha$ system.  With this observation and
the significant stretching of image \#1.1, we propose that the
Ly-$\alpha$ in the spectrum of image \#2.1 is scattered light from
Sextet Arc \#1.1, and that the redshift of image \#2.1 is detected by
{\it absorption} at $z$=2.534 toward the Sextet Arc \#1.1.  Note the
photometric redshift for image \#2.1 is $z=2.62 \pm 0.48$,
encompassing both redshifts $z$=2.534 and $z$=3.038 within the errors
(see Table~\ref{table3}).  If image \#2.1 does represent another
spatial region of Source \#1 at $z$=3.038, then these significant
variations in the Ly-$\alpha$ profile may have rather interesting
implications for the kinematics, as is discussed in \S6.

\section{Discussion}

\subsection{Ly-$\alpha$ Trends for the Sextet Arcs }
The Ly-$\alpha$ profiles in the spectra of images \#1.1, \#1.4 and
\#2.1 range from strong absorption to a combination of emission plus
absorption, to pure emission (Figure~\ref{fig_1d}).  The spectra for
images \#1.1 and \#1.4 appear to be representing two different spatial
regions of the same spatially-resolved LBG at $z$=3.038, and the
spectrum for image \#2.1 possibly does as well, depending on the
identification of its source of starlight.  The spectrum for image
\#1.1 was taken of a one arcsec portion of what is a giant arc with a
total length of more than 5.5 arcsec and an estimated area
magnification factor of $\gtrsim16$ (Figure~\ref{fig10}).  In turn,
image \#1.4, with a smaller estimated magnification of $\sim8$, has a
total spatial extent that is smaller than the slitwidth used to record
the spectrum, so that our data contains all the light from Source \#1.
Image \#2.1 has an estimated magnification of $\sim16$.  Its spectrum
shows a second and different one-arcsec portion of the giant arc
\#1.1, or a companion galaxy in a pair or group of galaxies at the
systemic redshift of the Sextet Arcs.

We can compare the equivalent widths of Ly-$\alpha$ for the three
images with the four LBG subsamples established by \citet{Shapley:03}.
By sorting their sample of 811 galaxies by rest equivalent width of
Ly-$\alpha$, W$_{Ly\alpha}$, they found a significant dependence of
W$_{Ly\alpha}$ on rest equivalent with of the strongest low ionization
interstellar absorption lines W$_{LIS}$.
For each image \#1.1, \#1.4, and \#2.1, we measure rest equivalent
width of Ly-$\alpha$, W$_{Ly\alpha}$, with the uncertainty determined
by continuum placement in the Lyman-$\alpha$ forest and the position
of the wavelength boundaries.  For image \#1.1, there are at least
four features that carve into the Ly-$\alpha$ absorption line profile
at $z$=3.038.  In the blue damping wing a second Ly-$\alpha$ line
appears at $z$=2.783.  At the systemic redshift there is a partial
emission-filling of the Ly-$\alpha$ line, and just redward of this
emission feature there are the two absorption features Si~IV
$\lambda$$\lambda$ 1393, 1402 at $z$=2.53.  We selected wavelength
boundaries by eye from 1182.0 - 1241.6 \AA.  For image \#1.4,
Ly-$\alpha$ is seen both in emission and absorption.  The equivalent
width is computed across both components, from 1193.0 - 1223.6 \AA.
For the metal lines, the total rest equivalent width of Si II
$\lambda$1260, O I$\lambda$1302+Si II$\lambda$1304, and C II
$\lambda$1334 was taken, with the errors based on the noise and
continuum placement.  Although the errors are large, particularly for
the lowest resolution data of images \#1.4 and \#2.1, W$_{LIS}$
decreases in strength as W$_{Ly\alpha}$ increases from $-26$ \AA \ in
absorption to 5.4 \AA \ in emission (Figure~\ref{fig11}).

These measurements suggest significant variations of Ly-$\alpha$
strength are possible within a {\it single} galaxy.  Note even if our
spectrum for image \#2.1 is not a spatial extended region of the
Sextet Arc \#1.1, there is still a large variation in both the
strength and profile of Ly-$\alpha$ for images \#1.1 and \#1.4.  The
equivalent widths from the literature are included for other SLLBGs,
where available.  For cB58, W$_{Ly\alpha}$ was not reported by
\citet{Pettini:02}.  It is visibly damped at their resolution of 58 km
s$^{-1}$, so in principle one could infer a value of W$_{Ly\alpha}$.
It is interesting that their value for W$_{LIS}$ is higher than the
Sextet Arcs and four LBG subsamples by a factor of two.

At least two of the three images are representing different parts of
an individual galaxy, thus fixing the LBG age.  Thus if the dependancy
of Ly-$\alpha$ strength with the strength of the metal lines is real,
then we infer that this trend does not appear to be a function of LBG
age.  To address whether the correlation may depend on metallicity,
\citet{Shapley:03} measured a small line ratio of Si II $\lambda$1260
and $\lambda$1527, and so do we for image \#1.1, implying that Si II
is saturated, and therefore that differences in equivalent width do
not depend on metallicity, but rather on the combination of covering
factor and a range of velocities over which the gas is absorbed.

\subsection{The Absorption System at $z$=2.53}

The absorption system at $z$=2.534 is seen toward two of our spectra
over a baseline of two arcsec on the sky.  Strong Ly-$\alpha$
absorption is seen toward both images \#1.1 and \#2.1.  While the data
are of insufficient resolution to measure column densities, there is
evidence in spectrum \#2.1 of absorption that clearly reaches the
zero-level of the continuum.  It is interesting that our spectrum
\#1.1 is at an estimated unlensed physical separation of 2$h^{-1}$ kpc
from the center of our slit for spectrum \#2.1, and yields Ly-$\alpha$
that is strong but probably unsaturated (Figure~\ref{fig10}).  It is
tempting to conclude that we are seeing the HI column density decrease
with radius over a size scale a factor of ten larger than the
half-light radius of image \#2.1.  However, two counterarcs of image
\#2.1 are predicted between images \#1.1 and \#1.2.  Thus it is
possible that our spectrum for image \#1.1 suffers from contaminated
light from these faint additional images.

The high ionization metal lines in this absorption system at $z$=2.534
are similar in strength to those in the Sextet Arc \#1.1 at $z$=3.038.
It is interesting that interstellar C~IV and Si~IV should be so
prominent in an {\it absorption} system at our low resolution.  We
consider briefly the possibility that this absorption system is not an
intervening object, but rather the spectrum of an outflow absorption
from the Sextet Arcs LBG.  The velocity difference between the
$z=3.038$ and $z=2.5$ systems is 38,000 km~s$^{-1}$.  Such a large
velocity difference for an associated outflow is not unusual for QSOs,
but the the Sextet Arc \#1.1 is an image of an LBG and does not show
any signs of AGN-like activity.

\subsection{The Lens Model}

Our lens model is taken from the surface mass density map in
\citet{Broadhurst:05}.  Briefly, the model was constructed based on 30
multiply-lensed galaxies, by minimizing the angular distance between
the predicted images and the observed images in the image plane.  The
model consists of the sum of a smoothly-varying low-frequency
component representing the dark matter, and a high-frequency component
representing the nonnegligable cluster galaxy contribution.  The
resulting mass map reproduces the positions and of the $106$ images
with a precision of typically 1-3\arcsec \ from the best fit model,
and accurately predicts the morphology, size, orientation, and parity
of the lensed images.  The critical curve in the region of Sextet Arcs
images \#1.1 and \#1.2, and Quintet Arcs image \#2.1, is shown in
Figure~\ref{fig10}.  The local cusp is produced by the massive nearby
cluster elliptical at $z$=0.187.  The region of high magnification
with positive parity (yellow-red), and high magnification with
negative parity (blue-white), are shown.  The thin black line at the
interface between these two regions marks locus of points for which
the magnification factor diverges.
This model has predictive power to estimate redshifts.
A comparison of the observed separations between the Source \#1 and \#2 image pairs 
\#1.3 and \#2.4, \#1.4 and \#2.2, and \#1.5 and \#2.3 
with model-predicted locations for different source redshifts 
suggests that source \#1 has a higher redshift.   Thus the model favors the scenario in which
Source \#1 and \#2 have redshifts $z$=3.038 and $z$=2.534 respectively.

If Sources \#1 and \#2 appear only as image pairs, there should be the
same number of total images of each source, but instead there are
seven images of the Sextet Arcs (but identified in
\citet{Broadhurst:05} as six separate ones), and five images of the
Quintet Arcs.  Two images of the Sextet Arcs appear as two merging
images, which is referred to as image \#1.1.  Their locations are
predicted by the model, at the intersection of the critical curve with
the giant arc (Figure~\ref{fig10}).  In a similar manner, two additional
unseparated images of Source \#2 are predicted at the position of the
second region of intersection of the critical curve with the giant arc
(between images \#1.1 and \#1.2).  These images are created from a
small portion of image \#2.1 which does not contain the bright
emission peak seen in the center of the 1 arcsec slit in
Figure~\ref{fig10}. This arc is possibly matched with a faint
stretched arc that is barely visible in the ACS image, between images
\#1.1 and \#1.2.  It is possible that these additional faint images
may have been detected by their {\it absorption}, even though our
spectrum for image \#1.1 falls $\sim1$ arcsec short of covering the
most likely position of these new predicted images.  Our spectrum of
image \#1.1 shows a strong intervening absorption system at $z$=2.534
which is also seen in our spectrum of image \#2.1.  However, this
absorption system is weaker in Ly-$\alpha$ than in its counterpart
absorption at the same redshift in our image \#2.1
(Figure~\ref{fig10}).  From this we conclude that this absorption
system may be the predicted unseparated counterarcs from a small
portion of image \#2.1.  Alternatively, if this intervening absorption
system forms a spatially-contiguous extension with origin at the
center of image \#2.1, we be may detecting the relatively unusual case
of a drop-off of H~I column density with radius at high-$z$.
Spectroscopy along the long axis of this giant arc will enable source
identifications of these absorption systems, thus yielding additional
constraints on the effort to derive a precision massmap for A1689.

\subsection{Intrinsic Properties of the Sextet Arcs}

The Sextet Arcs have an estimated reddening of $E(B-V) \sim 0.47,
0.10, 0.17$, for images \#1.1, \#1.4, and \#2.1, respectively.  As
discussed above, the extinction in image \#1.1 suffers from
contamination by a nearby cluster elliptical.  We can use these values
for the extinction plus the rest-frame ultraviolet flux to compute a
star formation rate.  Specifically, the ultraviolet flux of each best
SED fit is measured within a synthetic tophat filter of rest-frame
width 300 \AA, from 1250-1550 \AA.  Given the galaxy's redshift, we
convert $F_{\nu}(1400\textrm{\AA})$ for image \#1.1 to a luminosity
$L_{\nu}(1400\textrm{\AA})$, which in turn can be converted to a star
formation rate:

\begin{equation}
7.14\times10^{27}\frac{\textrm{ergs}}{\textrm{s}\cdot\textrm{Hz}}\label{eq:SFR}\rightarrow1\,\frac{M_{\odot}}{\textrm{yr}}\end{equation}

This conversion rate is computed for a Salpeter (1955) IMF truncated between $0.1\ <\ M/M_{\odot}\ <\ 100$.  Allowing the Salpeter IMF to extend from
$0.1~<~M/M_{\odot}~<125$ results in a conversion rate of
$8\times10^{27}$ ergs/s/Hz per $M_{\odot}$/year.  This is prefered by
some authors. Meanwhile a Scale (1986) IMF yields $3.5\times10^{27}$
ergs/s/Hz per $M_{\odot}$/year. If the currently fashionable Kroupa
(2001) IMF proves to be more accurate, then our SFR estimates should
be multiplied by $2/3$; the Kroupa conversion rate is
$1.07\times10^{28}$ ergs/s/Hz per $M_{\odot}$/year. The conversion we
adopt in Eq. \ref{eq:SFR} is employed by \citet{Hopkins:04}.  Objects
\#1.1, \#1.4, and \#2.1 have star formation rates of $\sim$40, 25, and
50 $M_{\odot}~{\rm yr}^{-1}$, respectively.  We correct these stellar
values by the extinction imposed by interstellar gas and dust as:
$E(B-V)_{\star} = 0.44 E(B-V)_{\rm gas}$. This attenuation in
magnitudes is: $A(\lambda) = E(B-V)_{\star} k(\lambda)$, where
$k(1400\textrm{\AA}) = 10.775$.  \citep{Calzetti:00}.  Upon
also correcting for the magnification, we measure star formation
rates (SFRs) of $\sim$4, $\sim$4, and $\sim$5 $M_{\odot}~{\rm
yr}^{-1}$ for images \#1.1, \#1.4, and \#2.1, respectively.  This SFR
is roughly one-tenth lower than that found for the general LBG
population of 25-52 $M_{\odot}~{\rm yr}^{-1}$ \citep{Shapley:03}.
However, it is similar to the SLLBG recently found at $z$=4.88, for
which a SFR of $12 \pm 2$ $M_{\odot}~{\rm yr}^{-1}$ has been measured
\citep{Swinbank:07}.

The Sextet Arcs are also rather faint, with unlensed apparent
magnitudes of $K$=$25.2 \pm 0.2$, $<27.4$, and $25.3 \pm 0.2$ for
\#1.1, \#1.4, and \#2.1 respectively.  The brightness for image \#1.1
is three magnitudes fainter in $K$ than $K^*$.  This makes the Sextet
Arcs quite faint, and not consistent with $L_V^*$ for the $z\sim3$
LBGs \citep{Shapley:01}.  Clearly more data taken at higher resolution
and along the long axis of \#1.1 in particular will be well rewarded
by providing spatially-resolved physical and kinematic information for
this rather unusual LBG.

 

   

\section{Summary and Future Work}

We have undertaken a spectroscopic redshift survey and present spectra
of the strongly-lensed LBG, the Sextet Arcs, at $z$=3.038.  The Sextet Arcs are
remarkable for their large apparent total magnitude of
$r_{625}=21.71$.  Our results are as follows:


\begin{enumerate}

\item{Spectra of the Sextet Arcs at $z_{sys}$=3.038 are presented for
two different images, \#1.1 and \#1.4.  The spectrum centered on image
\#2.1 is also presented, which is either a companion galaxy at the
same systemic redshift of the Sextet Arcs or, as we think more likely,
a different spatial region of image \#1.1.  This yields up to three
spatial regions across a single LBG: two different parts of image
\#1.1, and the whole of image \#1.4.}

\item{Unusually, Ly-$\alpha$ does not show evidence for gas motions.
The velocity difference between Ly-$\alpha$ and the low ionization
interstellar absorption lines is $\Delta v = -75 \pm 150$ km s$^{-1}$.
However, the line profile of Si~IV $\lambda$1394 shows broad,
blueshifted absorption indicative of gas outflow.}

\item{Across the spectra for the three images, \#1.1, \#1.4, and
\#2.1, the Ly-$\alpha$ profile changes dramatically, and W$_{IS}$
decreases in strength with increasing W$_{\alpha}$, similar to the
four LBG subsamples of \citet{Shapley:03}.  We emphasize that for the
Sextet Arcs this correlation all takes place within a {\it single}
galaxy.}

\item{Intrinsically, the Sextet Arcs have a rather modest SFR and
luminosity.  We infer SFR of $\sim4$, $\sim4$, and $\sim5$
$M_{\odot}~{\rm yr}^{-1}$ for images \#1.1, \#1.4, and \#2.1,
respectively.  This SFR is very different to that found for the
general LBG population, roughly one-tenth as high.  The Sextet Arcs
are also rather faint intrinsically, $K=25.2 \pm 0.13$ for image
\#1.1, which corresponds roughly to $\sim0.1K^*$.}

\item{The same intervening absorption system at $z$=2.534 is seen
towards our spectra of image \#1.1, where it is possibly damped, and
image \#2.1, which is weaker.  The angular separation is two arcsec on
the sky, or an unlensed physical size of $\sim2h^{-1}$kpc for an
estimated tangential stretch factor of four.  We interpret this either
as evidence of another counterimage of image \#2.1, or, if it is a
spatially-contiguous extension of image \#2.1, of a dropoff of H I
column density with radius from an origin centered on image \#2.1.  }

\item{We present a spectroscopic catalog in the field of A1689
comprising 98 secure redshifts, 82 of which are arclets, $z>$0.23 (see
Tables~\ref{table1} \& \ref{table3}).  This survey increases the
number of known arclets by six-fold.  We augment our catalog with
the spectroscopic redshifts of all known cluster members and measure a new cluster redshift for A1689
of $z$=0.187.}

\end{enumerate}

LBGs with $r_{625}$-band magnitudes brighter than 23 are still fairly
rare, but no longer mere oddities.  The newest member, the Sextet Arcs
at $z$=3.038 presented in this paper, is unique for showing a
spatially-resolved strong intervening absorption system as well.
While SLLBGs are anomalously bright, strongly-magnified, and
spatially-resolved due to gravitational lensing, how their intrinsic
characteristics compare with the general LBG population is still a
work in progress.  As more SLLBGs are discovered and studied, we will
be afforded valuable information on how intrinsic properties such as
the luminosity, extinction and SFR compare with the general LBG
population, and more generally, with the evolutionary state of
galaxies at $z \sim3$.


\acknowledgments We would like to thank Holland Ford, Alice Shapley, Bruce Draine,
and Todd Tripp for useful discussions, and Ray Murphy for his
technical assistance.  The ACS was developed under NASA contract NAS
5-32865.  BLF acknowledges support from Science Foundation Ireland
Research Frontiers Programme Grant PHY008.  DVB is funded through NASA
Long Term Space Astrophysics Grant NNG05GE26G.  The authors wish to
recognize and acknowledge the very significant cultural roleAand
reverence that the summit of Mauna Kea has always had within the
indigenous Hawaiian community.  We are most fortunate to have the
opportunity to conduct observations from this mountain.




Facilities: \facility{HST(ACS)}, \facility{Keck:I(LRIS)}, \facility{VLT}, \facility{Magellan:Baade ()}.

\appendix
\section{Spectroscopic Catalog}

\subsection{General Description}

A total of 98 objects with secure redshifts in the A1689 field are presented in Tables~\ref{table2} 
and \ref{table3}.  The columns are:  right ascension and declination, spectroscopic redshift, $I_{AB}$
magnitude, Bayesian photometric redshift, date of observation and telescope reference.  Only 
objects with secure redshifts appear in the tables, and that we take to mean 
 two or more spectroscopic features be detected 
at the 2$\sigma$ level in the continuum.  In the case of a single emission line, we require 
that it be detected blueward of rest-frame H$\alpha$ and have a photometric redshift consistent
with its identification.
All other objects with only one
spectroscopic feature are considered insecure and not included in this catalog.
The magnitude distribution of all of our objects with spectroscopic redshifts shows our catalog to
be complete down to roughly $I_{AB}\sim22.5$.

\begin{figure}
\includegraphics[viewport= 30 5 250 350,scale=1.2]{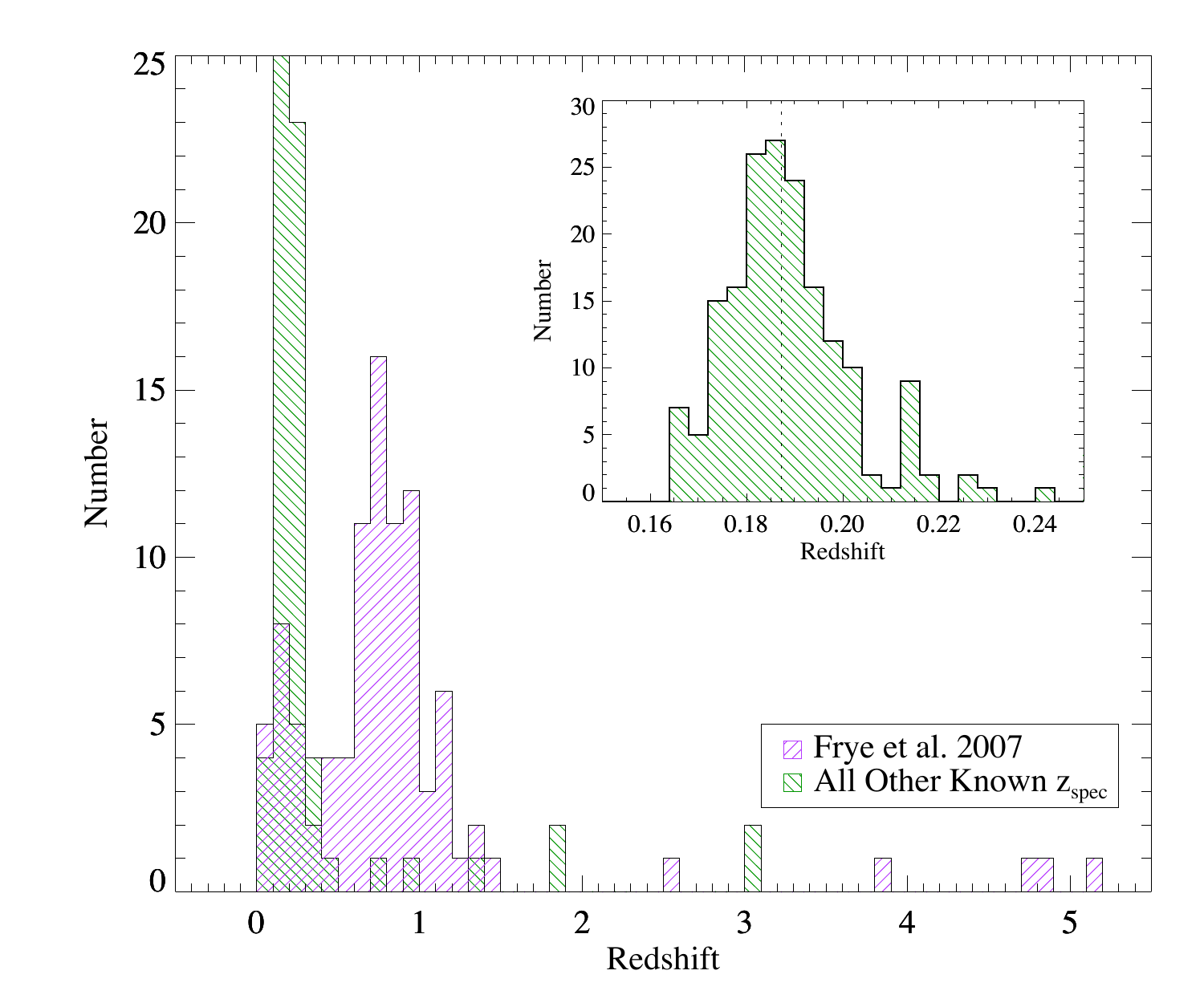}
\caption{Histogram of all published spectroscopic redshifts for Abell 1689.  The positive 
and negative slope
fill patterns indicate numbers of objects from the catalog in this paper, and those from all other 
published spectroscopic redshift catalogs respectively.
The criss-cross fill pattern shows the intersection between the catalogs.
The inset histogram shows the redshift distribution of the cluster members.
A new mean cluster redshift of $z$=0.187 is obtained, as indicated. Note the preponderance 
of background objects contributed by this paper.  This survey contributes
 72 new objects in the background, thus increasing the
number of arclets by sixfold. \label{fig4}}
\end{figure}

Our catalog is augmented by objects drawn from the literature.
This larger catalog of spectroscopic redshifts  is shown 
in 
Figure~\ref{fig4}.  The objects are
divided into two categories: those from this survey (positive slope
fill pattern) and all other published sources (negative slope fill
pattern) that are nonredundant \citep{Broadhurst:05,
Mieske:05, Frye:02, Duc:02, Balogh:02, Teague:90}.  Additionally, objects in common
between the catalogs were removed.  In the inset is shown the
redshift histogram for the cluster members from all
published sources, including this catalog.    Our catalog makes a significant
contribution to the number of arclets.  Of the 98 objects, 
82 are in the background, and 72 of those are new
spectroscopic redshifts.  Note our 10 duplicate arclets are included in
Tables~\ref{table1} and \ref{table3} as they are the result of new data and are useful for the
purposes of confirmation.  For A1689, there are 186 cluster members in total in the
`master' catalog, with ten new contributions out of the 11 cluster members in our sample.
From these 186 cluster members we calculate a new mean cluster
redshift of $z$=0.1872.  Finally, five objects in our catalog are stars and are
included in this paper for completeness.

The arcs at $z$=3.038 were  first published in \citet{Broadhurst:05}, and their
references are revised from Table~1 as follows:
the spectroscopic redshifts for images \#1.1 and \#1.4 were first referenced in \citet{Broadhurst:05}, and not in
\citet{Frye:02} and Fort et~al.~(1997), as given in their Table 1. 
As for other catalogs, one object from the \citet{Teague:90}
survey was removed as its position lies between two objects,
(RA, DEC) = (13:11:30.516, -01:20:45.82).  Also the positions of
four objects from \citet{Mieske:05} are revised from what appears in their paper as follows:
Candidate 2 (RA, DEC) = (13:11:33.11, -01:19:22.5),
Candidate 3 (RA, DEC) = (13:11:29.90, -01:20:05.5),
Candidate 4 (RA, DEC) = (13:11:31.09, -01:21:42.1), and
Candidate 5 (RA, DEC) = (13:11:25.73, -01:21:14.8). All
coordinates are J2000.


\subsection{Multiply-imaged Systems}

\begin{figure}
\includegraphics[viewport= -40 70 150 600, scale=0.85]{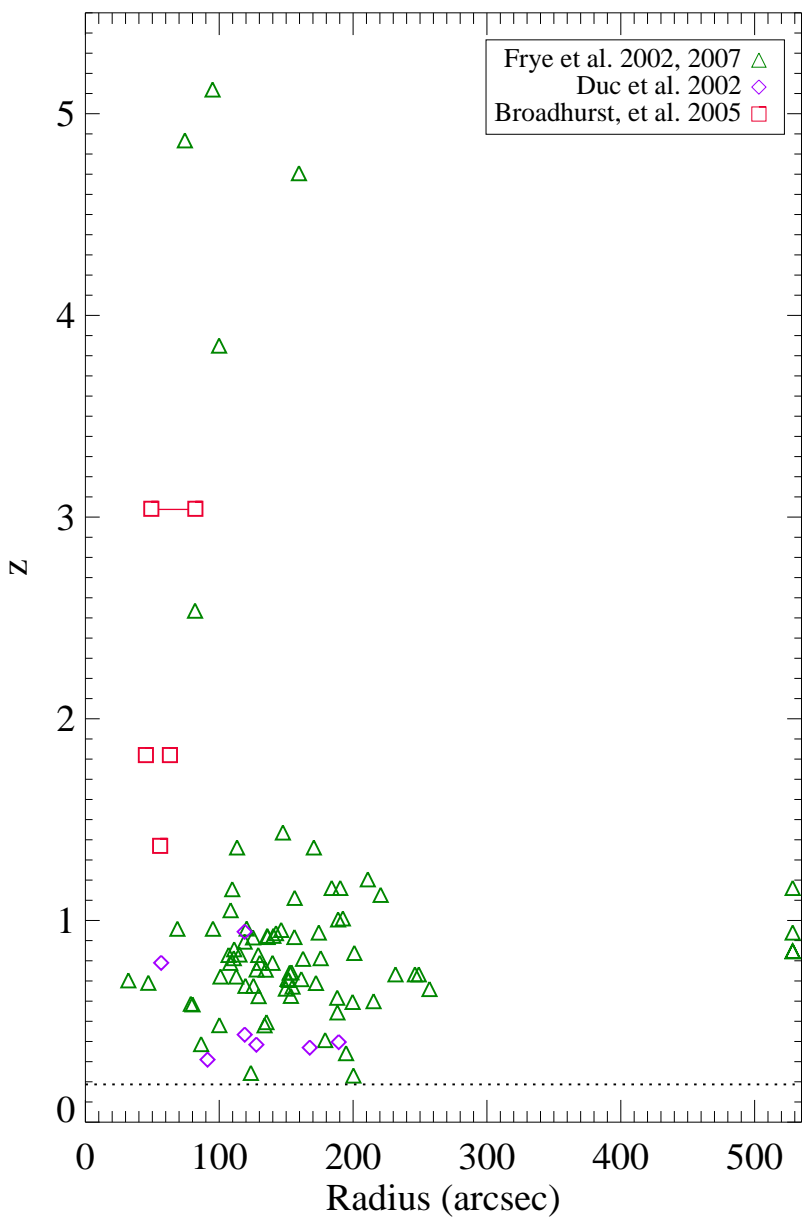}
\caption{Spectroscopic redshift as a function of radius for all objects 
with published redshifts $z>$0.23 in the field of A1689.
The cluster redshift of $z$=0.187 is indicated by the dotted line,
and the different symbols give the contributing references.  Six galaxies have been
discovered with $z>2.5$.  Note the
two points at $z$=3.038 are two images of the same galaxy, the Sextet Arcs, and so
are counted as one object.
\label{fig5}}
\end{figure}

All spectroscopically-confirmed $z>0.23$ objects in the field of 
Abell 1689 are shown in Figure~\ref{fig5}.  The symbols
denote different source references, as indicated.  This survey 
increases the number of known  
background objects significantly, from 12 to 72.
Six objects are clearly separated from the
bulk of faint background galaxies to be at high-z ($z>2.5$) and small in radius.  
Note there are five high-z objects and six high-z points. 
The two points at $z$=3.038 represent two images of the Sextet Arcs,  which is the focus of this paper.
Four images of multiply-imaged systems are in our spectroscopic sample, Sources \#1.1 and \#1.4 (of the Sextet Arcs),
Source \#2.1 (one of the Quintet Arcs) and Source \#7.1.
  These images were introduced in
\citet{Broadhurst:05}, in which 30 multiply-imaged systems were identified.

The highest spectroscopically-confirmed  
redshifts in the A1689 field come from emission line objects at $z=4.868$ (\#7.1) and at $z=5.12$ 
\citep{Frye:02}. Object \#7.1 is triply-imaged, and FORS2 VLT data show it to have significant structure in the Lyman-series forest.
The data include the region below the Lyman-limit, and will be
presented in an upcoming paper. The galaxy at $z=4.71$ is new, and was
also discovered with FORS2 at the VLT.


\subsection{Photometric Redshifts}

\begin{figure}
\includegraphics[viewport=10 5 5 400,scale=1.]{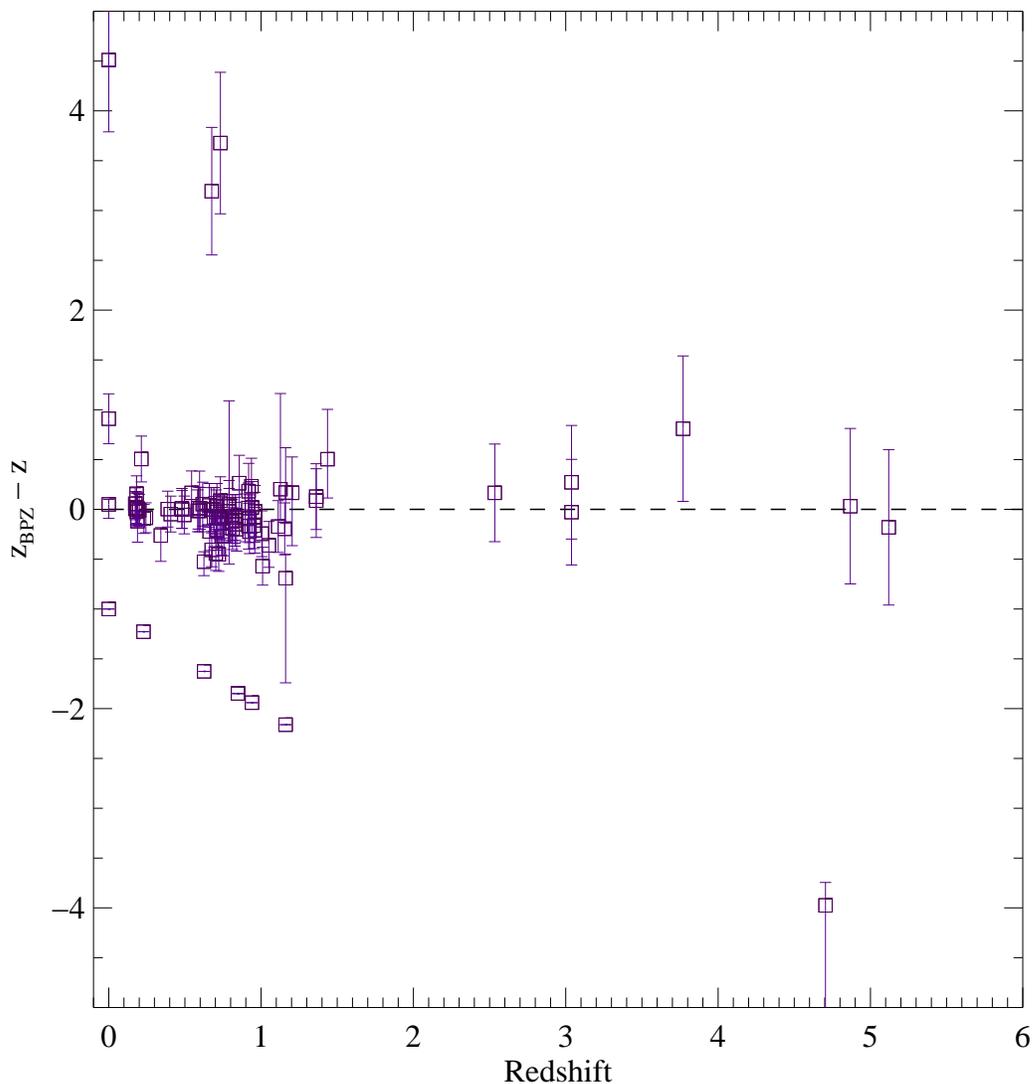}
\caption{Bayesian photometric redshifts,  $z_{BPZ}$, are compared with spectroscopic redshifts from our sample.     When the four outliers are removed,
we find the best fit BPZ redshifts agree with the spectroscopic redshifts
to within $z_{BPZ} = 0.11 (1 + z_{spec})$.  The photometry was meticulously computed based on galaxy-subtracted images to
correct for significant contamination from cluster members.
\label{fig6}}
\end{figure}

\begin{figure}
\includegraphics[viewport=10 13 500 500, scale=0.6]{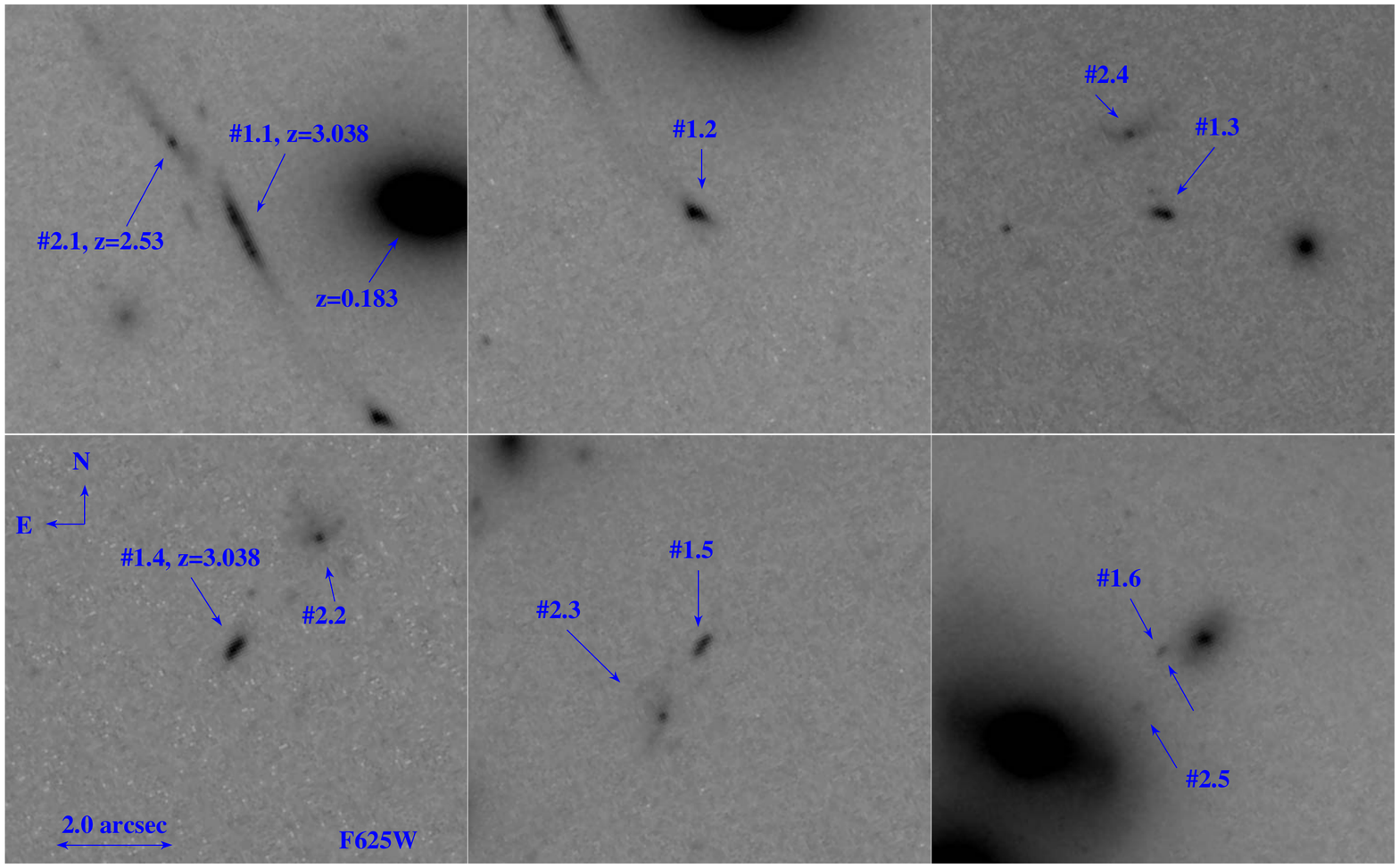}
\caption{All six separate images of the Sextet Arcs at $z$=3.038 (Source \#1) are shown,
including labels and spectroscopic redshifts, if known.  
Images \#1.1 and \#1.2 comprise a giant arc that is spatially-extended on the sky, 
$>5.5$ arcsec.   Source \#2 is also multiply-imaged, and all five of this set of images, the Quintet
Arcs, are also marked.   Note the similar morphologies of the image pairs of 
sources \#1 and \#2 from panel to panel
despite being stretched, rotated, and tangentially and radially parity-flipped.
\label{fig7}}
\end{figure}

\begin{figure}
\includegraphics[viewport=115 100 600 500,scale=0.69]{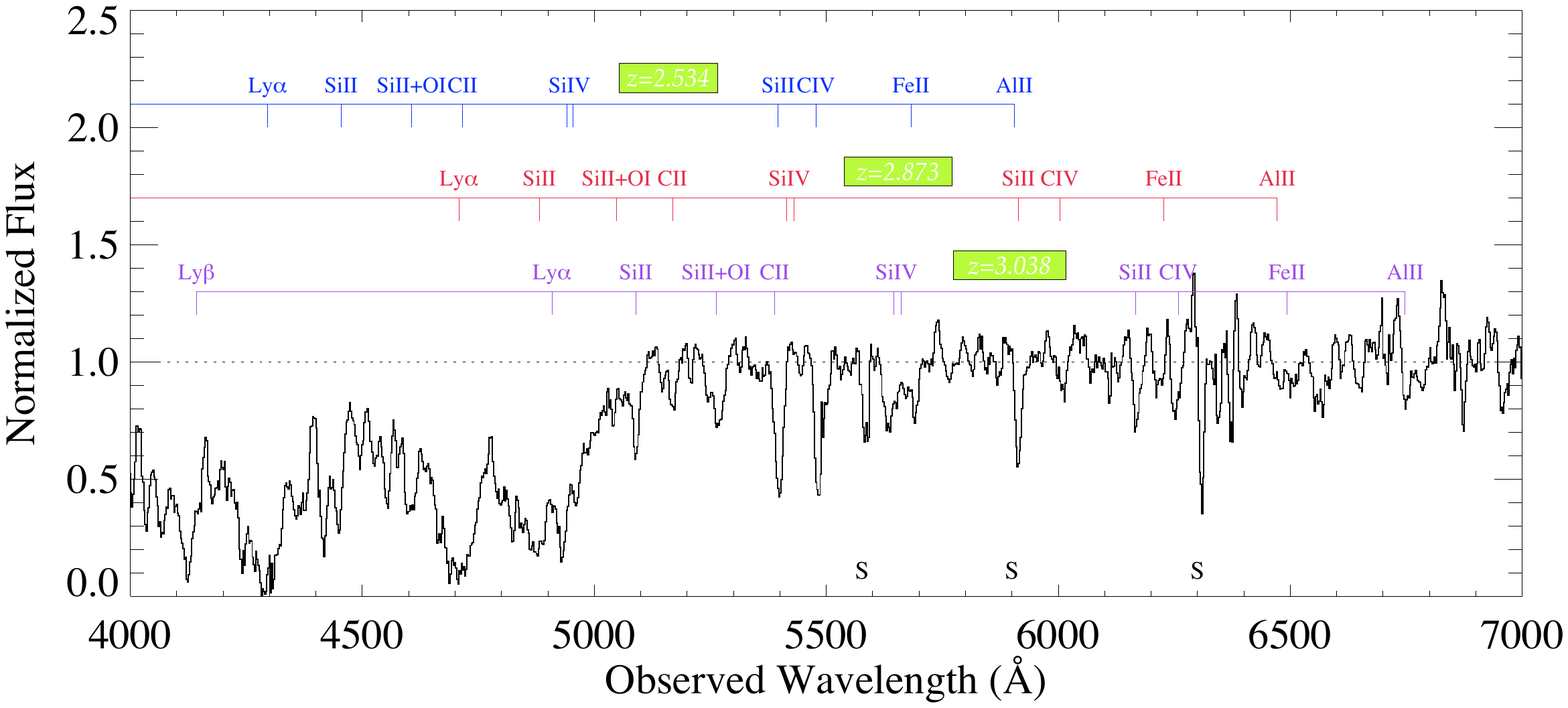}
\caption{Spectrum of the Sextet Arc \#1.1 of at $z$=3.038 is plotted vs. Observed Wavelength.  
There are many prominent interstellar absorption features, and two intervening absorption systems,
at $z$=2.873 and $z$=2.534.  The latter absorption system is spatially-extended on the sky, 
and shows stronger Ly-$\alpha$ absorption in our spectrum of image \#2.1 taken two arcsec away on the sky,
or 2$h^{-1}$kpc at $z$=2.534.  The positions of the prominent skylines in the spectrum are indicated.
 \label{fig_dvb}}
\end{figure}

\begin{figure}
\includegraphics[viewport=25 80 200 650,scale=0.85]{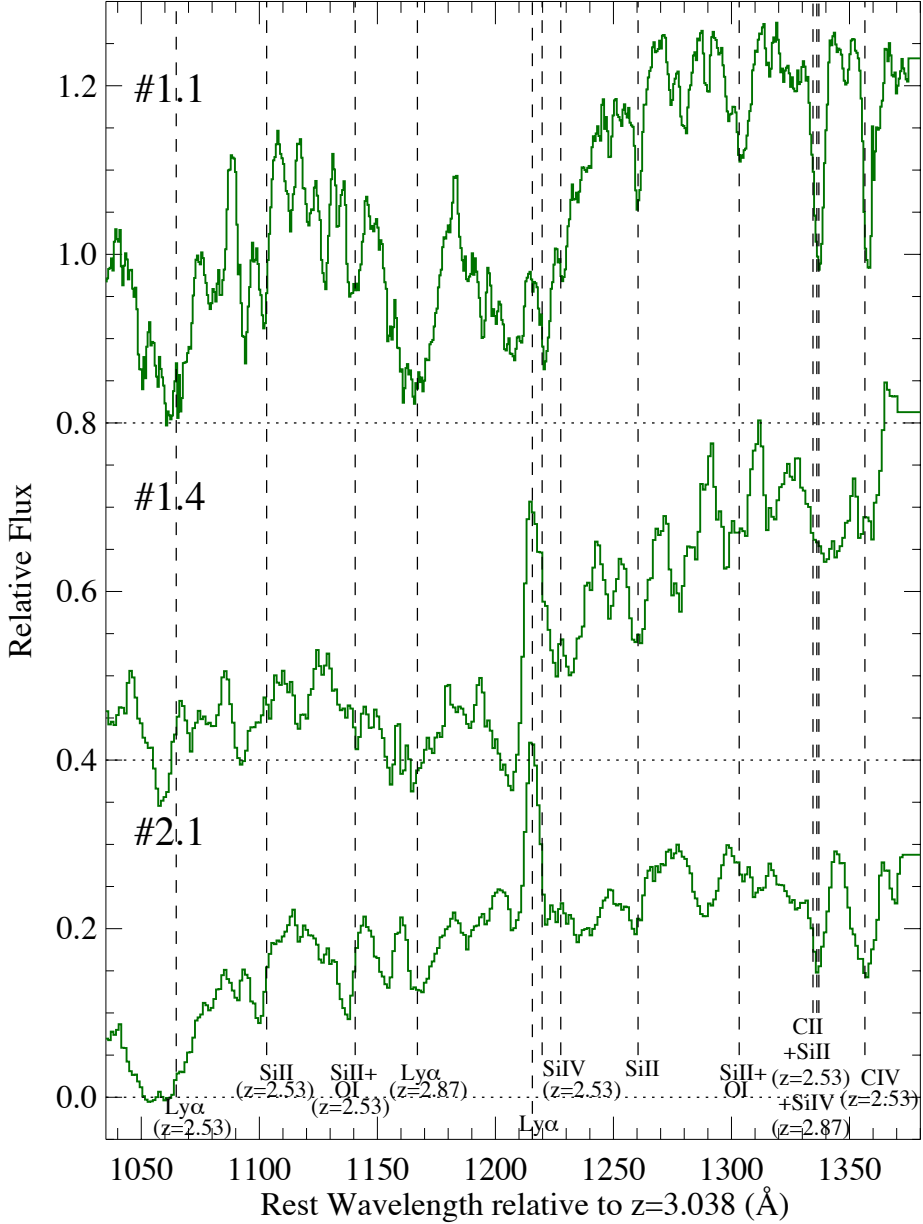}
\caption{Spectra taken of two different images of the sextuply-lensed galaxy at $z$=3.038, the Sextet Arcs.     The high signal-to-noise spectrum for image \#1.1 is rich in features, and corresponds
to a one arcsec spatial region of this giant arc $>5.5$ arcsec in extent.  
The spectrum for \#1.4, by contrast, 
includes all of the light from this smaller arclet with lower magnification.
 A third spectrum is also shown, centered on image \#2.1, and taken at a 
 two arcsec separation from our spectrum for image \#1.1.
The Ly-$\alpha$ profile between all three spectra 
shows considerable variations, from
strong absorption, 
 to a combination of absorption plus emission and strong emission.
 \label{fig_1d}}
\end{figure}

\begin{figure}
\includegraphics[viewport=15 -20 300 350,scale=0.87]{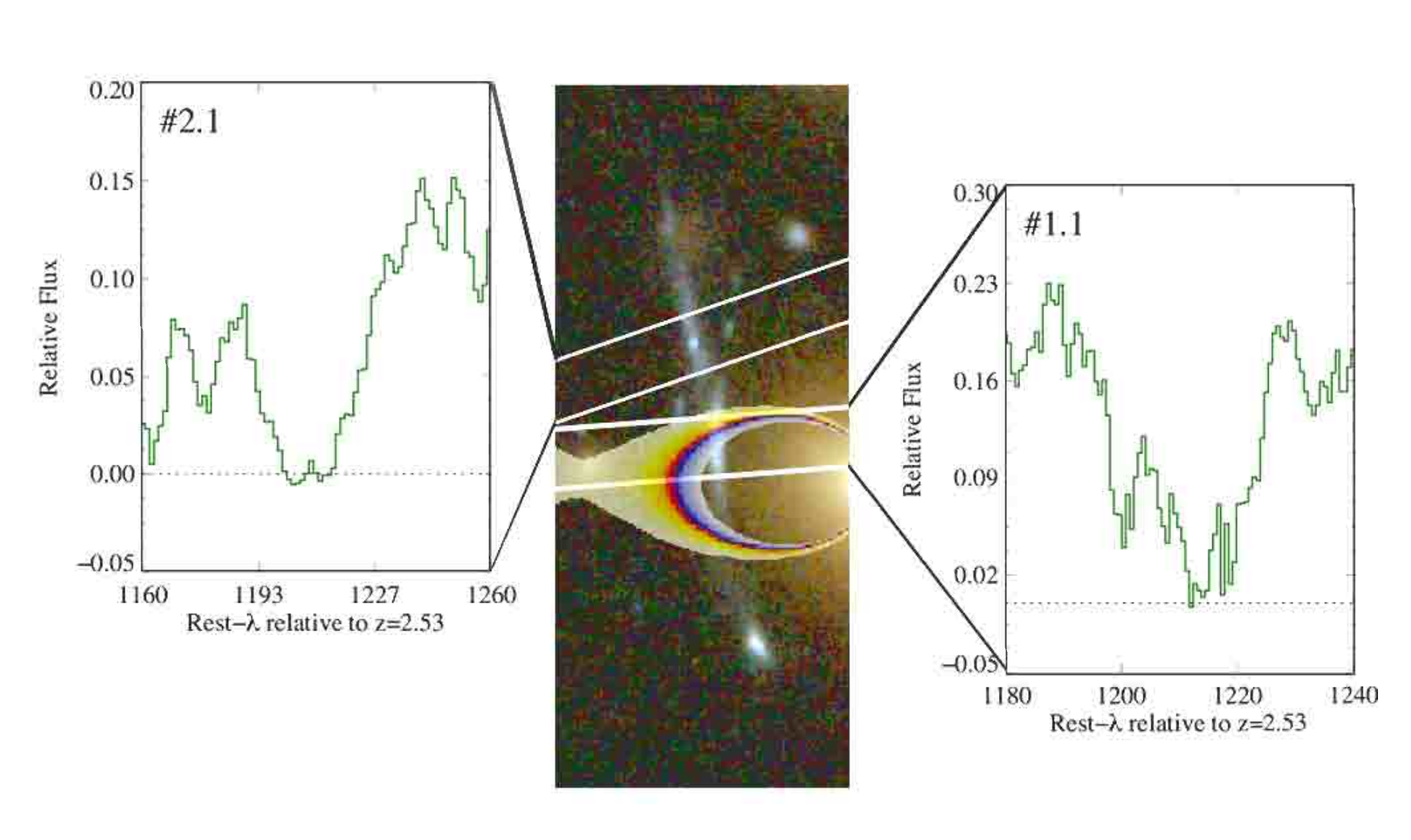}
\caption{Ly-$\alpha$ profiles for the strong intervening absorption system at $z$=2.534.
The observational setup is overlayed, showing our two 1\arcsec \ slits at two different spatial
positions of this intervening LBG. In the spectrum centered on image \#2.1,  Ly-$\alpha$
at $z$=2.534 
is strongly absorbed and reaches the bottom of the continuum.
By contrast, in our spectrum of image \#1.1 of the Sextet Arcs, there is also 
significant absorption by Ly-$\alpha$, but it is not obviously saturated.  
The critical curve is overlayed, showing the region of high magnification with
positive parity (yellow-red) and high magnification with negative parity (blue-white). 
The intersections of the critical curve with this giant arc mark the predicted positions of  
a fold arc of the Sextet Arcs, which is seen, and a fold arc of the Quintet Arcs,
which is predicted to be fainter and if present, may be detected only by {\it absorption} 
towards the Sextet Arcs. We conclude that  if the spatially-resolved absorption system 
seen at $z$=2.535 is a spatially-contiguous extension of image \#2.1, we are detecting a drop
off of H I column density with radius, and if not, we may be detecting additional counterimages
of Source \#2.  \label{fig10}
}  
\end{figure}

\begin{figure}
\includegraphics[viewport=60 70 200 450,scale=1.1]{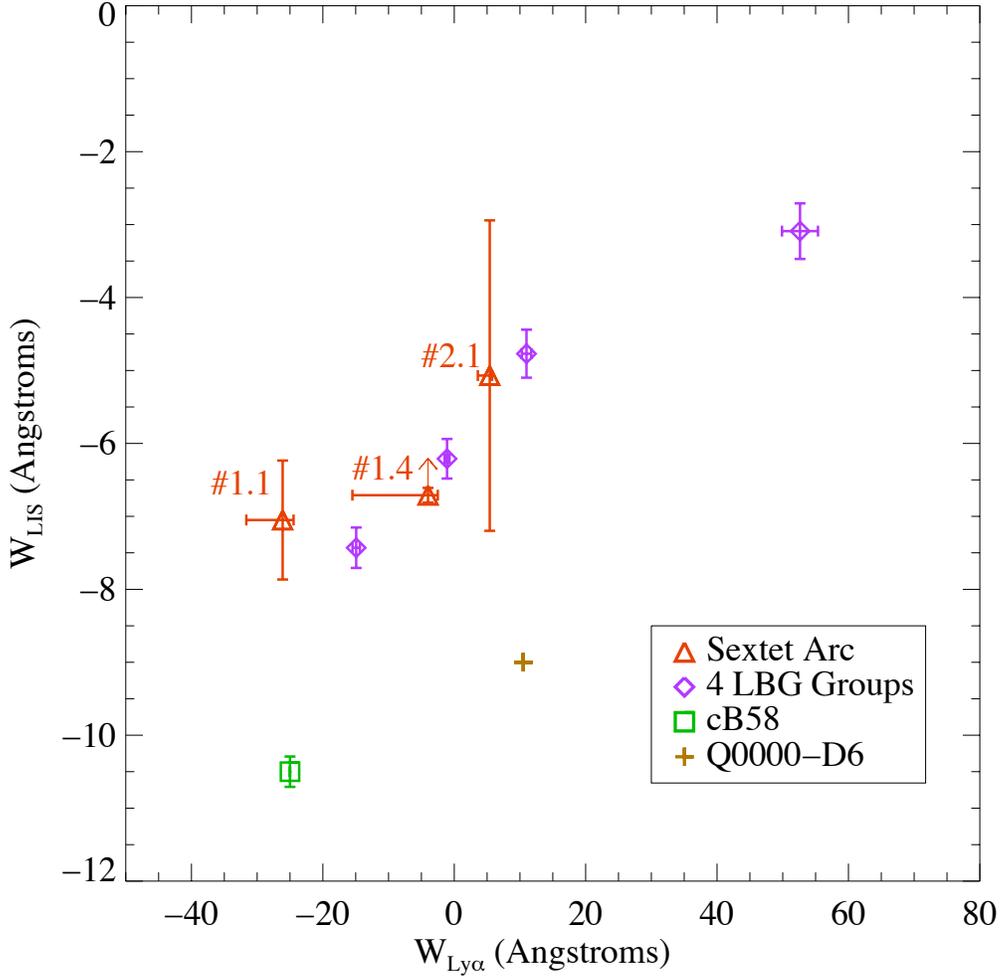}
\caption{The total rest equivalent width of the strong low ionization interstellar absorption
lines (W$_{LIS}$) is plotted against the rest equivalent width for Ly-$\alpha$.  The three triangular-shaped points correspond to each of our three spectra:  image \#1.1, image \#1.4,
and image \#2.1.  There is a trend emerging of decreasing W$_{LIS}$ with increasing W$_{Ly\alpha}$.
The diamond-shaped points correspond to the four LBG subsamples
of \citet{Shapley:03}.  Also included are
cB58 \citep{Pettini:02}, and Q0000-D6 \citep{Giallongo:02}.  
  This trend is similar to that found for the four LBG subsamples but
in our case are seen across a {\it single} galaxy.
\label{fig11}}

\end{figure}

The photometric redshifts given in Tables~\ref{table1} and~\ref{table3} are derived using a Bayesian approach.
The Bayesian photometric redshift (BPZ) method is described in detail in other papers
\citep{Benitez:04,Benitez:00},
and is found to meet the requirements of a field that is faint, crowded, and magnified.
The comparison of spectroscopic and photometric redshifts is shown in Figure~\ref{fig6}.
There are two clear outliers at $z_{spec} \sim 0.7$, one object with similarly large deviance
but with 95\% error bars that encompass the true redshift of $z_{spec} = 4.71$, and one
object at $z=0$.
When these three objects are removed,
we find the best fit BPZ redshifts agree with the spectroscopic redshifts
to within $0.11 (1 + z_{spec})$.

This scatter is somewhat larger than that achieved in previous BPZ studies, with the most
recent example being the Ultra Deep Field (UDF) \citep{Coe:06}.
The main uncertainty in the photometric redshift calculations is traced back to
the difficulty of determining reliable magnitudes for faint objects in rich cluster fields in comparison
to other field studies such as the UDF or Great Observatories Origins Deep Survey (GOODS) \citep{Mobasher:04}.
Our sample here is characterized by lensed objects with very low surface brightness.  
The images are background-limited and in many cases, extended into giant arcs, 
leading to larger photometric errors than comparable unlensed galaxies to similar
depth, such as GOODS.
Another source of uncertainty is the crowded field.
Light from many of our sample galaxies is contaminated by light from cluster members.
We have done our best to subtract this cluster light to improve the photometry (\S3.2),
but the galaxy-modeling is not perfect.
Apart from the unavoidable 'root-n' noise, the subtraction introduces both random error 
and an error that is filter-dependent and most prominent for the fainter objects.

Additional improvements in the photometric redshifts should result from
use of the \citet{Coe:06} BPZ template set, which includes younger (bluer) starbursts.
And the reported agreement between photometric and spectroscopic redshifts should tighten further
once galaxies with poor SED fits (i.e., $\chi^2_{mod} < 1$) are removed from the comparison.
These issues will all be addressed with the release of the full A1689 photometric redshift catalog
(Coe et al., in prep.).




\bibliography{bfrye,apj-jour.bib}







\clearpage

\begin{deluxetable}{cccc}
\tabletypesize{\scriptsize}
\tablecaption{Log of A1689 Spectroscopic Observations}
\tablewidth{0pt}
\tablehead{ Site & Date & Exp (h) & Grating (lines/mm) }
\startdata
Keck LRIS                   & Apr 1996 (1)  & 2        & 300 at 5000 \AA  \\    
$^{\prime\prime}$      & Apr 1996 (2)  & 2        &$^{\prime\prime}$ \\  
$^{\prime\prime}$      & Jun 1998        & 2.8     & $^{\prime\prime}$ \\ 
$^{\prime\prime}$      & Mar 1999       & 2.2      & 400 at 8500 \AA   \\ 
$^{\prime\prime}$      & Apr 1999        & 2         & $^{\prime\prime}$ \\ 
$^{\prime\prime}$      & Mar 2003 (1) & 1.9      & $^{\prime\prime}$ \\  
 $^{\prime\prime}$     & Mar 2003 (2) & 1.9      & $^{\prime\prime}$ \\ 
VLT FORS2                & Jun/Jul 2001 & 11       & 300 at 8600 \AA  \\ 
Magellan LDSS-2     & May 2003       &$2-14$& 300 at 5500  \AA  \\ 
 \enddata
\label{table2}
\end{deluxetable}

\clearpage

\begin{deluxetable}{ccccl}
\tabletypesize{\scriptsize}
\tablecaption{Spectroscopic Features for the Spatially-resolved 
Sextet Arc at $z$=3.038}
\tablewidth{0pt}
\tablehead{
\colhead{Source \#} & \colhead{Identification} & \colhead{$\lambda_{obs} 
$} & \colhead{W$_0$} \\
    & & \colhead{(\AA)} & \colhead{(\AA)} & \colhead{Comments}
}
\startdata
\#1.1 & HI & 1215.67 & $-26_{+1.7}^{5.4}$  & \\

& Si II & 1260.42 & $-1.9\pm0.42$ &  \\
& SiII+OI & 1303.27\tablenotemark{a} & $-2.2\pm0.61$& \\
& C II & 1334.53 & $-3.0\pm0.36$ & Blend:  SiIV $\lambda$1394 ($z$=2.87) + SiII $\lambda$1527 ($z$=2.53) \\
& SiIV & 1393.76 & $-2.1\pm0.42$ & \\
& SiIV & 1402.77 & $-1.2\pm0.34$& Blend with FeII $\lambda$1608 ($z$=2.53)\\ 
& SiII & 1526.71 & $-4.1\pm1.28$ & \\
& CIV & 1549.48\tablenotemark{b} & $-0.88\pm0.30$&  \\
& FeII & 1608.45 & $-1.1\pm0.57$& \\
& AlII & 1670.79 & $-1.2\pm0.62$& \\

\#1.4 & HI &1215.67 & -4.0 $_{-1.5}^{+11.5}$&   \\
          &SiII, SiII+OI, CII &1260.42,1303.27,1334.53 & -6.7$_{+2.1}^{-1.1}$ &  \\
    &SiII & 1526.71 & -2.0$_{+0.22}^{-1.0}$ & \\

\#2.1  & HI & 1215.67 &4.0$^{+1.5}_{-5.0}$ & \\
          &SiII, SiII+OI, CII &1260.42,1303.27,1334.53 & $-5.1\pm2.1$  & \\

\enddata
\tablenotetext{a}{This is the mean vacuum wavelength of O I $\lambda$ 
1302 and Si II $\lambda$ 1304, and
all quantities measured refer to the total of these two lines which 
are blended at our spectral resolution. }

\tablenotetext{b}{This is the mean vacuum wavelength of the C IV $\lambda$1548, 1550 doublet,
which is blended in this spectrum.}

\label{table4}
\end{deluxetable}

\clearpage

\begin{deluxetable}{ccllcll}
\tabletypesize{\scriptsize}
\tablecaption{Spectroscopic Identification of $z<$2.5 Objects towards A1689}
\tablewidth{0pt}
\tablehead{
\colhead{RA} & \colhead{DEC} & \colhead{$z$} & \colhead{{\it $I_{AB}$}} & \colhead{$z_{BPZ}$} & \colhead{Date} & \colhead{Telescope}
}
\startdata
13:11:19.289&-01:19:43.42& 0.228& 22.83$\pm$ 0.46 &---                     & 96 Apr & Keck LRIS     \\ 
13:11:20.231&-01:20:01.10& 0.81 & 22.51$\pm$ 0.03 & 0.69$^{+0.22}_{-0.22}$ & 03 May & Magellan      \\ 
13:11:20.346&-01:21:15.94& 0.626& 21.90$\pm$ 0.45 &---                     & 96 Apr & Keck LRIS     \\ 
13:11:21.056&-01:17:31.19& 0.733& 21.41$\pm$ 0.01 & 0.65$^{+0.22}_{-0.22}$ & 03 May & Magellan      \\ 
13:11:21.503&-01:17:47.40& 0.733& 22.80$\pm$ 0.03 & 0.82$^{+0.24}_{-0.24}$ & 03 May & Magellan     \\ 
13:11:21.853&-01:17:26.80& 0.733& 23.14$\pm$ 0.04 & 4.41$^{+0.71}_{-0.71}$ & 03 May & Magellan     \\ 
13:11:22.523&-01:20:39.43& 0.960& 23.03$\pm$ 0.02 & 0.94$^{+0.26}_{-0.26}$ & 99 Mar & Keck LRIS     \\ 
13:11:22.781&-01:19:01.65& 0.709& 24.06$\pm$ 0.06 & 0.50$^{+0.20}_{-0.20}$ & 96 Apr & Keck LRIS    \\ 
13:11:22.843&-01:17:06.32& 0.66 & 22.33$\pm$ 0.02 & 0.70$^{+0.22}_{-0.22}$ & 03 May & Magellan     \\ 
13:11:24.104&-01:18:52.65& 0.672& 21.96$\pm$ 0.01 & 0.67$^{+0.22}_{-0.22}$ & 03 May & Magellan       \\ 
13:11:24.155&-01:19:56.54& 1.155& 21.34$\pm$ 0.01 & 0.96$^{+0.26}_{-0.26}$ & 98 Jun &  Keck LRIS    \\ 
13:11:24.240&-01:19:52.68& 0.857& 21.91$\pm$ 0.02 & 1.12$^{+0.28}_{-0.28}$ & 98 Jun & Keck LRIS     \\ 
13:11:24.367&-01:19:36.87& 0.895& 23.75$\pm$ 0.06 & 0.81$^{+0.24}_{-0.24}$ & 98 Jun & Keck LRIS     \\ 
13:11:24.609&-01:19:20.83& 0.757& 24.16$\pm$ 0.09 & 0.69$^{+0.22}_{-0.22}$ & 03 Mar & Keck LRIS     \\ 
13:11:24.652&-01:20:03.38& 0.481& 21.33$\pm$ 0.01 & 0.48$^{+0.19}_{-0.19}$ & 96 Apr &  Keck LRIS   \\ 
13:11:24.802&-01:20:23.08& 0.0  & 21.84$\pm$ 0.01 & 5.27$^{+0.82}_{-0.82}$ & 96 Apr &   Keck LRIS  \\ 
13:11:24.960&-01:19:36.61& 0.722& 24.23$\pm$ 0.11 & 0.27$^{+0.17}_{-0.17}$ & 03 Mar &  Keck LRIS   \\ 
13:11:25.617&-01:18:01.21& 1.005& 22.61$\pm$ 0.03 & 0.76$^{+0.23}_{-0.23}$ & 03 May & Magellan      \\ 
13:11:26.237&-01:19:56.45& 0.183& 18.10$\pm$ 0.01 & 0.34$^{+0.18}_{-0.18}$ & 98 Jun & Keck LRIS    \\ 
13:11:26.714&-01:19:37.49& 0.959& 21.0 $\pm$ 0.42 & 0.75$^{+0.23}_{-0.23}$ & 99 Mar &  Keck LRIS   \\
13:11:26.919&-01:20:00.65& 0.0  & 22.35$\pm$ 0.03 & 0.05$^{+0.14}_{-0.05}$ & 03 May & Magellan      \\ 
13:11:27.122&-01:18:48.43& 0.184& 19.84$\pm$ 0.01 & 0.16$^{+0.15}_{-0.15}$ & 03 May &  Magellan     \\ 
13:11:27.173&-01:18:49.98& 0.480& 21.5 $\pm$ 0.44 & 0.49$^{+0.20}_{-0.20}$ & 03 May &  Magellan     \\ 
13:11:27.191&-01:18:26.55& 1.112& 22.09$\pm$ 0.02 & 0.94$^{+0.26}_{-0.26}$ & 98 Jun &  Keck LRIS    \\ 
13:11:27.851&-01:20:07.65& 0.175& 17.48$\pm$ 0.01 & 0.18$^{+0.16}_{-0.16}$ & 03 May &  Magellan     \\ 
13:11:28.221&-01:20:50.93& 0.703& 22.00$\pm$ 0.02 & 0.26$^{+0.17}_{-0.17}$ & 03 Mar &  Keck LRIS    \\ 
13:11:28.325&-01:18:27.50& 0.710& 20.59$\pm$ 0.01 & 0.74$^{+0.23}_{-0.23}$ & 03 May &  Magellan     \\ 
13:11:28.547&-01:23:02.85& 0.756& 23.76$\pm$ 0.10 & 0.49$^{+0.20}_{-0.20}$ & 99 Mar &  Keck LRIS    \\ 
13:11:28.685&-01:17:37.54& 0.231& 22.38$\pm$ 0.02 & 0.14$^{+0.15}_{-0.14}$ & 01 Jun &  VLT FORS2    \\ 
13:11:28.886&-01:20:01.94& 0.180& 21.05$\pm$ 0.01 & 0.20$^{+0.16}_{-0.16}$ & 96 Apr &  Keck LRIS    \\ 
13:11:29.100&-01:19:46.93& 0.188& 19.31$\pm$ 0.01 & 0.20$^{+0.16}_{-0.16}$ & 03 Mar & Keck LRIS     \\ 
13:11:30.231&-01:22:46.10& 0.0  & 22.68$\pm$ 0.03 & 4.51$^{+0.72}_{-0.72}$ & 99 Mar &  Keck LRIS    \\ 
13:11:30.508&-01:19:34.67& 0.174& 20.66$\pm$ 0.02 & 0.23$^{+0.16}_{-0.16}$ & 03 Mar &  Keck LRIS    \\ 
13:11:30.677&-01:18:55.50& 0.676& 23.85$\pm$ 0.05 & 3.87$^{+0.64}_{-0.64}$ & 03 Mar &  Keck LRIS    \\ 
13:11:30.751&-01:21:38.78& 0.691& 20.32$\pm$ 0.01 & 0.61$^{+0.21}_{-0.21}$ & 99 Mar &   Keck LRIS   \\ 
13:11:31.472&-01:21:05.94& 0.189& 20.62$\pm$ 0.01 & 0.19$^{+0.16}_{-0.16}$ & 96 Apr &   Keck LRIS   \\ 
13:11:31.622&-01:23:21.20& 0.705& 21.23$\pm$ 0.01 & 0.68$^{+0.22}_{-0.22}$ & 03 May &  Magellan     \\ 
13:11:31.824&-01:17:49.29& 0.617& 21.12$\pm$ 0.01 & 0.67$^{+0.22}_{-0.22}$ & 01 Jul &  VLT FORS2    \\ 
13:11:32.025&-01:21:55.40& 0.959& 21.10$\pm$ 0.01 & 0.88$^{+0.25}_{-0.25}$ & 98 Jun &  Keck LRIS    \\ 
13:11:32.424&-01:24:18.19& 1.204& 24.17$\pm$ 0.08 & 1.37$^{+0.53}_{-0.36}$ & 99 Apr & Keck LRIS     \\ 
13:11:33.230&-01:19:16.95& 0.200& 20.37$\pm$ 0.01 & 0.19$^{+0.16}_{-0.16}$ & 98 Jun & Keck LRIS     \\ 
13:11:33.028&-01:19:14.64& 0.790& 22.98$\pm$ 0.07 & 0.82$^{+0.58}_{-1.06}$ & 96 Apr & Keck LRIS     \\ 
13:11:33.555&-01:19:01.50& 0.244& 22.11$\pm$ 0.02 & 0.16$^{+0.15}_{-0.15}$ & 96 Apr &Keck LRIS      \\ 
13:11:33.632&-01:22:02.00& 0.387& 22.92$\pm$ 0.04 & 0.39$^{+0.18}_{-0.18}$ & 03 May &Magellan       \\ 
13:11:33.950&-01:19:15.75& 1.362& 22.48$\pm$ 0.03 & 1.49$^{+0.33}_{-0.33}$ & 98 Jun &Keck LRIS      \\ 
13:11:34.211&-01:19:23.98& 1.051& 22.44$\pm$ 0.04 & 0.69$^{+0.22}_{-0.24}$ & 98 Jun &Keck LRIS      \\ 
13:11:34.318&-01:19:04.93& 0.676& 22.60$\pm$ 0.02 & 0.27$^{+0.17}_{-0.17}$ & 03 May &Magellan       \\ 
13:11:34.900&-01:18:35.62& 0.918& 22.31$\pm$ 0.02 & 1.10$^{+0.28}_{-0.28}$ & 96 Apr &Keck LRIS      \\ 
13:11:35.057&-01:21:26.00& 0.584& 21.22$\pm$ 0.01 & 0.57$^{+0.21}_{-0.21}$ & 98 Jun &Keck LRIS      \\ 
13:11:35.227&-01:20:30.16& 0.587& 20.64$\pm$ 0.01 & 0.60$^{+0.21}_{-0.21}$ & 98 Jun &Keck LRIS      \\ 
13:11:35.264&-01:19:01.50& 0.918& 22.27$\pm$ 0.03 & 0.75$^{+0.23}_{-0.23}$ & 96 Apr & Keck LRIS     \\ 
13:11:35.619&-01:21:52.59& 0.722& 23.56$\pm$ 0.04 & 0.65$^{+0.22}_{-0.22}$ & 03 Mar &Keck LRIS      \\ 
13:11:36.052&-01:19:24.71& 0.916& 22.85$\pm$ 0.03 & 0.93$^{+0.25}_{-0.25}$ & 99 Apr &Keck LRIS      \\ 
13:11:36.370&-01:19:06.25& 0.937& 23.66$\pm$ 0.05 & 1.17$^{+0.29}_{-0.28}$ & 96 Apr &Keck LRIS      \\ 
13:11:36.536&-01:19:25.03& 0.790& 20.50$\pm$ 0.01 & 0.77$^{+0.23}_{-0.23}$ & 96 Apr &Keck LRIS      \\ 
13:11:36.638&-01:22:32.72& 0.790& 24.17$\pm$ 0.10 & 0.84$^{+0.24}_{-0.24}$ & 98 Jun &Keck LRIS      \\ 
13:11:37.089&-01:19:26.10& 0.924& 21.60$\pm$ 0.01 & 0.70$^{+0.22}_{-0.22}$ & 96 Apr &Keck LRIS      \\ 
13:11:37.183&-01:21:40.60& 0.813& 21.78$\pm$ 0.01 & 0.73$^{+0.23}_{-0.23}$ & 96 Apr & Keck LRIS     \\ 
13:11:37.284&-01:21:06.02& 0.829& 23.91$\pm$ 0.04 & 0.68$^{+0.22}_{-0.22}$ & 03 Mar &Keck LRIS      \\ 
13:11:37.577&-01:21:24.34& 0.831& 22.02$\pm$ 0.01 & 0.75$^{+0.23}_{-0.23}$ & 98 Jun & Keck LRIS     \\ 
13:11:37.590&-01:23:00.70& 1.362& 24.11$\pm$ 0.07 & 1.45$^{+0.37}_{-0.32}$ & 03 Mar &Keck LRIS      \\ 
13:11:37.694&-01:19:49.79& 0.625& 20.95$\pm$ 0.01 & 0.10$^{+0.14}_{-0.10}$ & 03 May &Magellan*      \\ 
13:11:38.051&-01:19:58.25& 0.189& 22.18$\pm$ 0.01 & 0.07$^{+0.21}_{-0.07}$ & 96 Apr &Keck LRIS      \\ 
13:11:38.239&-01:21:42.05& 0.214& 21.98$\pm$ 0.02 & 0.72$^{+0.23}_{-0.23}$ & 03 May &Magellan       \\ 
13:11:38.348&-01:22:22.76& 0.741& 24.00$\pm$ 0.10 & 0.47$^{+0.19}_{-0.19}$ & 98 Jun & Keck LRIS     \\ 
13:11:38.653&-01:21:38.87& 0.496& 22.06$\pm$ 0.02 & 0.44$^{+0.19}_{-0.19}$ & 03 May & Magellan      \\ 
13:11:38.690&-01:22:17.78& 0.743& 21.77$\pm$ 0.02 & 0.65$^{+0.22}_{-0.22}$ & 03 May & Magellan      \\ 
13:11:38.791&-01:20:52.17& 0.829& 21.03$\pm$ 0.01 & 0.72$^{+0.23}_{-0.23}$ & 96 Apr & Keck LRIS     \\ 
13:11:38.846&-01:23:41.89& 0.600& 20.21$\pm$ 0.01 & 0.61$^{+0.21}_{-0.21}$ & 03 Mar & Keck LRIS     \\ 
13:11:39.101&-01:23:45.53& 1.127& 23.59$\pm$ 0.06 & 1.33$^{+0.31}_{-0.96}$ & 03 Mar &Keck LRIS      \\ 
13:11:39.472&-01:22:51.09& 1.161& 24.21$\pm$ 0.08 & 0.47$^{+1.05}_{-0.24}$ & 03 Mar &Keck LRIS      \\ 
13:11:39.524&-01:20:46.21& 0.924& 22.35$\pm$ 0.02 & 0.83$^{+0.24}_{-0.24}$ & 96 Apr &Keck LRIS      \\ 
13:11:39.544&-01:20:13.37& 0.953& 24.22$\pm$ 0.06 & 0.79$^{+0.23}_{-0.24}$ & 96 Apr &Keck LRIS      \\ 
13:11:39.748&-01:22:55.89& 1.161& 23.88$\pm$ 0.06 & 1.33$^{+0.31}_{-0.45}$ & 03 Mar &Keck LRIS      \\ 
13:11:39.804&-01:21:31.52& 0.662& 24.19$\pm$ 0.06 & 0.44$^{+0.19}_{-0.19}$ & 03 Mar & Keck LRIS     \\ 
13:11:39.833&-01:22:36.47& 0.407& 24.26$\pm$ 0.06 & 0.36$^{+0.18}_{-0.18}$ & 98 Jun & Keck LRIS     \\ 
13:11:39.891&-01:20:31.08& 1.436& 23.36$\pm$ 0.03 & 1.94$^{+0.39}_{-0.50}$ & 03 Mar &Keck LRIS      \\ 
13:11:40.204&-01:18:52.38& 0.0  & 21.50$\pm$ 0.01 & 0.91$^{+0.25}_{-0.25}$ & 96 Apr & Keck LRIS     \\ 
13:11:40.284&-01:23:02.45& 0.839& 23.91$\pm$ 0.07 & 0.64$^{+0.22}_{-0.21}$ & 03 Mar &Keck LRIS      \\ 
13:11:40.614&-01:19:38.24& 0.94 & 22.09$\pm$ 0.02 & 0.96$^{+0.26}_{-0.26}$ & 03 May & Magellan      \\ 
13:11:41.322&-01:21:44.92& 0.813& 22.71$\pm$ 0.06 & 0.62$^{+0.21}_{-0.21}$ & 96 Apr &Keck LRIS      \\ 
13:11:41.341&-01:22:39.09& 0.596& 23.91$\pm$ 0.12 & 0.58$^{+0.21}_{-0.40}$ & 03 Mar & Keck LRIS     \\ 
13:11:41.620&-01:20:54.24& 0.690& 20.35$\pm$ 0.01 & 0.69$^{+0.22}_{-0.22}$ & 03 May & Magellan      \\ 
13:11:41.986&-01:19:34.47& 0.342& 22.91$\pm$ 0.03 & 0.08$^{+0.26}_{-0.08}$ & 96 Apr & Keck LRIS     \\ 
13:11:42.154&-01:19:34.06& 1.01 & 19.35$\pm$ 0.01 & 0.44$^{+0.19}_{-0.19}$ & 96 Apr & Keck LRIS     \\ 
13:11:42.301&-01:20:07.79& 0.544& 21.66$\pm$ 0.01 & 0.71$^{+0.23}_{-0.22}$ & 03 May & Magellan      \\ 
13:11:42.838&-01:20:25.94& 0.848& 22.63$\pm$ 0.46 & ---                    & 96 Apr & Keck LRIS     \cr 
13:11:43.607&-01:20:36.93& 0.848& 23.26$\pm$ 0.46 & ---                    & 96 Apr &Keck LRIS      \cr 
13:11:43.825&-01:19:38.76& 0.940& 22.85$\pm$ 0.44 & ---                    & 96 Apr &Keck LRIS      \cr 
13:11:44.466&-01:20:30.58& 1.161& 22.72$\pm$ 0.46 & ---                    & 96 Apr &Keck LRIS      \cr 
13:11:46.056&-01:20:51.94& 0.0    & ---                            & ---                    & 96 Apr & Keck LRIS     \cr 
\enddata
\label{table1}
\end{deluxetable}

\begin{deluxetable}{cccccccc}
\tabletypesize{\scriptsize}
\tablecaption{$z>$2.5 Objects in the A1689 Field}
\tablewidth{0pt}
\tablehead{RA (J2000) & Dec (J2000) &Name\tablenotemark{a} & $z_{spec}$ 
& $z_{BPZ}$ & $g_{475}$ & $r_{625}$& $i_{775}$ }
\startdata
13:11:26.450& -01:19:56.75 & Sextet$-$1.1        &$3.038\pm 0.003$\tablenotemark{b}&3.03
$\pm$ 0.53&23.34 $\pm$ 0.01&22.55 $\pm$ 0.01&22.40 $\pm$ 0.01 \\
13:11:26.281& -01:20:00.26 & $^{\prime\prime}-$1.2 &$-$                 
 &3.04 $\pm$ 0.53&24.33 $\pm$ 0.01&23.60 $\pm$ 0.01&23.51 $\pm$0.01  \\
13:11:29.777& -01:21:07.48 & $^{\prime\prime}-$1.3 &$-$                 
 &3.27 $\pm$ 0.56&25.23 $\pm$ 0.02&24.53 $\pm$ 0.02&24.48 $\pm$0.02  \\
13:11:33.063& -01:20:27.40 & $^{\prime\prime}-$1.4 &$3.038\pm 0.003$\tablenotemark{c}&2.94
$\pm$ 0.52&24.67 $\pm$ 0.13&24.03 $\pm$ 0.01&24.02 $\pm$ 0.01  \\
13:11:31.935& -01:20:06.00 & $^{\prime\prime}-$1.5 &$-$                 
 &3.35 $\pm$ 0.57&25.46 $\pm$ 0.03&24.63 $\pm$ 0.02&24.56 $\pm$ 0.02  \\
13:11:29.853& -01:20:38.41 & $^{\prime\prime}-$1.6 & $-$                
  &1.06$^{+1.97}_{-0.27}$&25.02 $\pm$ 0.05&24.18 $\pm$ 0.05&23.94 $\pm$ 0.06
  \\
13:11:26.523& -01:19:55.45 & Quintet$-$2.1                               &2.534\tablenotemark{d}&2.62
$\pm$ 0.48&23.56 $\pm$ 0.01&23.27 $\pm$0.01 &23.07 $\pm$0.01 \\
 13:11:32.783 &-01:20:27.60 & $^{\prime\prime}-$2.2 & $-$ &2.54 & 24.03 $\pm$0.01
 & 23.85 $\pm$ 0.01 & 23.76 $\pm$0.01 \\
 13:11:31.776 &-01:20:09.24 &$^{\prime\prime}-$2.3  & $-$ &2.54 & 24.29 $\pm$0.02
 & 24.06 $\pm$ 0.02 & 24.10 $\pm$0.02 \\
 13:11:29.616 &-01:21:07.92 &$^{\prime\prime}-$2.4  & $-$ &2.54 & 24.42 $\pm$0.02
 & 24.23 $\pm$  0.02 & 24.14 $\pm$  0.02 \\
13:11:29.689  &-01:20:41.28 &$^{\prime\prime}-$ 2.5 & $-$ &2.54 & 24.26 $\pm$0.03
 & 23.45 $\pm$ 0.03  & 23.04 $\pm$ 0.03 \\
13:11:29.938& -01:19:14.65 & & 3.770\tablenotemark{b}&4.58 $\pm$ 0.73& $>27.2$
       &24.98 $\pm$ 0.12&24.13 $\pm$ 0.90  \\
13:11:24.066& -01:18:47.17 & & 4.705\tablenotemark{c}&0.73$^{+4.65}_{-0.23}$&
$-$            &$-$             &$-$    \\
13:11:25.449& -01:20:51.84 & \#7.1                 &4.868\tablenotemark{b}&4.92
$\pm$ 0.78&26.70 $\pm$ 0.67&25.12 $\pm$ 0.12&23.48 $\pm$ 0.03   \\
13:11:34.991& -01:19:51.07 & & 5.120\tablenotemark{b}&4.94 $\pm$ 0.78&$>27.2$
        &26.59 $\pm$ 0.70&25.58 $\pm$ 0.18   \\
\enddata

\tablenotetext{a}{Published name from \citet{Broadhurst:05}, where available}
\tablenotetext{b}{\citet{Broadhurst:05}}
\tablenotetext{c}{\citet{Frye:02}}
\tablenotetext{d}{This paper}
\label{table3}
\end{deluxetable}

\clearpage



\clearpage



\clearpage

\end{document}